\documentclass{aa}  

\usepackage{color}
\usepackage{graphicx}
\usepackage{amsmath}
\usepackage{mathrsfs}
\usepackage{txfonts}
\usepackage{natbib,twoopt}
\usepackage[breaklinks=true]{hyperref} %% to avoid \citeads line fills
\bibpunct{(}{)}{;}{a}{}{,}             %% natbib format for A&A and ApJ
\makeatletter
  \newcommandtwoopt{\citeads}[3][][]{\href{http://adsabs.harvard.edu/abs/#3}%
    {\def\hyper@linkstart##1##2{}%
     \let\hyper@linkend\@empty\citealp[#1][#2]{#3}}}
  \newcommandtwoopt{\citepads}[3][][]{\href{http://adsabs.harvard.edu/abs/#3}%
    {\def\hyper@linkstart##1##2{}%
     \let\hyper@linkend\@empty\citep[#1][#2]{#3}}}
  \newcommandtwoopt{\citetads}[3][][]{\href{http://adsabs.harvard.edu/abs/#3}%
    {\def\hyper@linkstart##1##2{}%
     \let\hyper@linkend\@empty\citet[#1][#2]{#3}}}
  \newcommandtwoopt{\citeyearads}[3][][]%
    {\href{http://adsabs.harvard.edu/abs/#3}
    {\def\hyper@linkstart##1##2{}%
     \let\hyper@linkend\@empty\citeyear[#1][#2]{#3}}}
\makeatother
\definecolor{mygreen}{RGB}{0, 139, 69}

\DeclareSymbolFont{mymath}{T1}{ybv}{m}{it}
\SetSymbolFont{mymath}{normal}{T1}{ybv}{m}{it}
\DeclareSymbolFontAlphabet{\mathnormal}{mymath}
\DeclareMathSymbol{\vel}{\mathalpha}{mymath}{`v}

\begin{document} 

   \title{On the variation of carbon abundance in galaxies\\ 
     and its implications}

%   \subtitle{}

   \author{D. Romano
          \inst{1},
%          \fnmsep\thanks{donatella.romano@inaf.it},
          M. Franchini
          \inst{2},
          V. Grisoni
          \inst{3,4,2},
          E. Spitoni
          \inst{5},
          F. Matteucci
          \inst{4,2,6}
          and C. Morossi
          \inst{2}
          }

   \institute{INAF, Astrophysics and Space Science Observatory, 
     Via Gobetti 93/3, 40129 Bologna, Italy\\
     \email{donatella.romano@inaf.it}
     \and INAF, Osservatorio Astronomico di Trieste, 
     Via Tiepolo 11, 34131 Trieste, Italy
     \and SISSA, Via Bonomea 265, 34136 Trieste, Italy
     \and Dipartimento di Fisica, Sezione di Astronomia, 
     Universit\`a di Trieste, Via Tiepolo 11, 34131 Trieste, Italy
     \and Stellar Astrophysics Centre, Department of Physics and Astronomy, 
     Aarhus University, Ny Munkegade 120, 8000 Aarhus C, Denmark
     \and INFN, Sezione di Trieste, Via Valerio 2, 34127 Trieste, Italy
             \\
%             \thanks{The university of heaven temporarily does not
%                     accept e-mails}
             }

   \date{Received 17 March 2020 / Accepted 11 May 2020}

   \titlerunning{Carbon abundance in galaxies}
   \authorrunning{Romano et al.}

   \abstract
      { The trends of chemical abundances and abundance ratios observed in 
        stars of different ages, kinematics, and metallicities bear the 
        imprints of several physical processes that concur to shape the host 
        galaxy properties. By inspecting these trends, we get precious 
        information on stellar nucleosynthesis, the stellar mass spectrum, the 
        timescale of structure formation, the efficiency of star formation, as 
        well as any inward or outward flows of gas. In this paper, we analyse 
        recent determinations of carbon-to-iron and carbon-to-oxygen abundance 
        ratios in different environments (the Milky Way and elliptical 
        galaxies) using our latest chemical evolution models that implement 
        up-to-date stellar yields and rely on the tight constraints provided by 
        asteroseismic stellar ages (whenever available). A scenario where most 
        carbon is produced by rotating massive stars, with yields largely 
        dependent on the metallicity of the parent proto-star clouds, allows us 
        to fit simultaneously the high-quality data available for the local 
        Galactic components (thick and thin discs) and for microlensed dwarf 
        stars in the Galactic bulge, as well as the abundance ratios inferred 
        for massive elliptical galaxies. Yet, more efforts are needed from both 
        observers and theoreticians in order to base these conclusions on 
        firmer grounds.}

   \maketitle
%
%-------------------------------------------------------------------

   \section{Introduction}
   \label{sec:intro}

   Carbon is one of the most abundant elements in the Universe and a powerful 
   key to understanding stellar and galactic structure and evolution.

   In stellar interiors, carbon is an important contributor to the opacity 
   \citep[e.g.,][and references therein]{2002A&A...387..507M} and, as a 
   catalyst in the CNO cycle, it affects the energy generation that accompanies 
   thermonuclear fusion \citep{1939PhRv...55..434B,1957RvMP...29..547B}. The 
   ratio of the abundances of the two stable isotopes of carbon, $^{12}$C and 
   $^{13}$C, measured at the surface of stars evolved off the main sequence and 
   compared to evolutionary model predictions, informs us about the occurrence 
   and strength of nuclear burning and mixing processes along the giant 
   branches \citep{1994A&A...282..811C,2000A&A...354..169G,2019A&A...621A..24L,
     2019ApJ...878...43M}. On the other hand, the anticorrelation between C and 
   N abundances observed in main-sequence globular cluster (GC) stars cannot 
   be attributed to internal stellar processes \citep{1998MNRAS.298..601C,
     2003AJ....125..197H}; rather, it adds up to the many peculiarities of 
   stellar populations in GCs \citep[see][for a recent 
     review]{2019A&ARv..27....8G} that still wait for a satisfactory 
   explanation \citep{2018ARA&A..56...83B}. Carbon detections also prove useful 
   for the characterisation of the atmospheres of exoplanets 
   \citep[e.g.,][]{2004ApJ...604L..69V}.

   On the scale of galaxies, carbon abundances can be used to constrain the 
   timescales of structure formation \citep[e.g.,][]{2012ceg..book.....M} and 
   to pin down the shape of the galaxy-wide stellar initial mass function 
   \citep[gwIMF,][]{2018A&A...620A..39J} in objects where direct estimates are 
   impossible \citep{2017MNRAS.470..401R,2018Natur.558..260Z}. Carbon also 
   traces the distribution of cool gas in local and distant galaxies via 
   molecular and atomic fine structure line emission, thus providing the 
   necessary link between stellar and star formation rate densities in the 
   Universe \citep{2013ARA&A..51..105C}. Carbon monoxide further traces giant 
   molecular outflows in bright quasar hosts and starburst-dominated galaxies 
   \citep[e.g.,][]{2010A&A...518L.155F,2014A&A...562A..21C}, shedding light on 
   the ultimate driver of such gigantic outward motions of processed gas.

   From the incomplete list of topics above it is already apparent that carbon 
   is an element of paramount importance in many research fields of 
   contemporary astrophysics. In particular, a firm understanding of the origin 
   and evolution of carbon in the Milky Way remains a crucial step towards the 
   development of theoretical models aimed at explaining C abundance 
   measurements in other systems: indeed, such models should always be tested 
   and comply with local data first. In this respect, it is worth recalling 
   that the stellar yields of carbon (i.e., the amounts of newly-produced C 
   that dying stars eject in their surroundings) are not firmly established yet 
   and, therefore, the fractional contribution to the carbon abundance from 
   different sources is still under debate \citep[see, e.g.,][for a recent 
     reappraisal of this problem]{2017MNRAS.470..401R,2019MNRAS.490.2838R}.

   Nowadays, the flood of data secured by large modern spectroscopic surveys, 
   such as the Gaia-ESO Public Spectroscopic Survey 
   \citep[GES,][]{2012Msngr.147...25G}, the Apache Point Observatory Galactic 
   Evolution Experiment \citep[APOGEE,][]{2017AJ....154...94M}, and the 
   GALactic Archaeology with HERMES \citep[GALAH,][]{2015MNRAS.449.2604D}, 
   provides the raw material for sophisticated analyses involving thousands of 
   stars that belong to different Galactic components. As a result, 
   statistically significant trends of abundance ratios with metallicity can be 
   obtained for different stellar populations, which can then be meaningfully 
   compared to the predictions of theoretical models.

   In this paper we deal with the origin and evolution of carbon in galaxies. 
   First, we compare the results of our Galactic chemical evolution (GCE) 
   models to the [C/Fe] versus [Fe/H] and [C/O] versus [O/H] abundance ratios 
   obtained by \citet{2019A&A...630A.104A} from a reanalysis of literature data 
   for a sample of nearby late-type stars including departures from local 
   thermodynamic equilibrium (LTE) and corrections for three-dimensional (3D) 
   stellar atmospheres. The theoretical results are confronted also with median 
   sequences of [C/X] versus [X/H] (where X~= Fe, O and/or Mg) derived for two 
   larger samples of solar neighbourhood stars, one selected from the GALAH 
   second data release \citep[DR2,][]{2019ApJ...886...84G} and the other from 
   the GES fifth internal data release \citep[iDR5,][]{2020ApJ...888...55F}. 
   The different data sets are further compared to each other (see next 
   section). Second, we comment on [C/X] determinations in other environments, 
   focusing on the evolution of carbon at high metallicities. Lastly, basing on 
   the comparison between model predictions and observations, we discuss the 
   fractional contribution to C enrichment that is coming from different 
   stellar polluters.

   The paper is organised as follows. In Sect.~\ref{sec:data} we describe the 
   data used for the comparison with the outputs of the models that, in turn, 
   are portrayed in Sect.~\ref{sec:models}. Sect.~\ref{sec:results} presents 
   our results, which are further discussed in Sect.~\ref{sec:disc}. Finally, 
   Sect.~\ref{sec:conc} reports our conclusions.

   \section{Observational data}
   \label{sec:data}

%-------------------------------------- Two column figure
   \begin{figure*}[ht]
     \centering
     \includegraphics[width=0.48\textwidth]{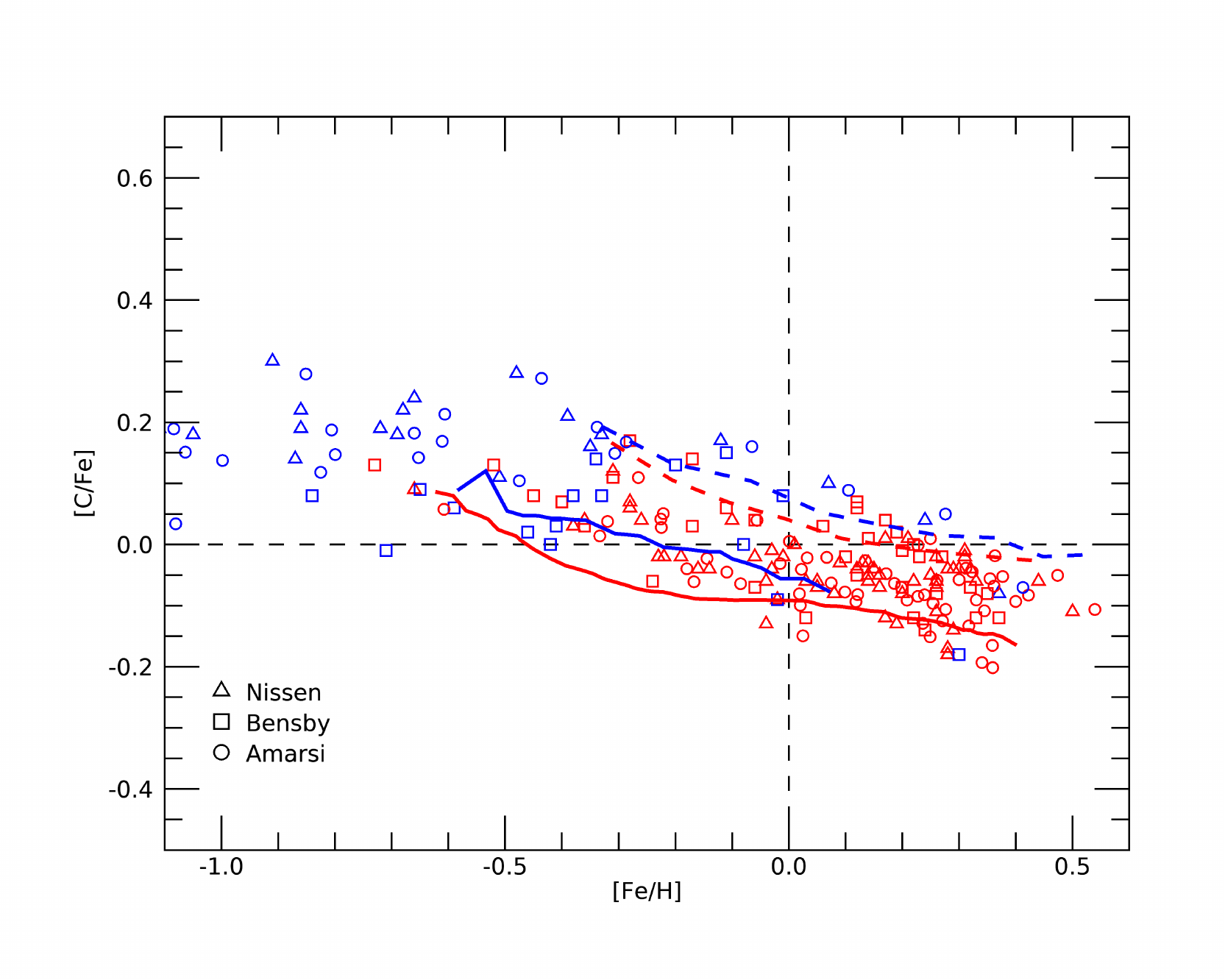}
     \includegraphics[width=0.48\textwidth]{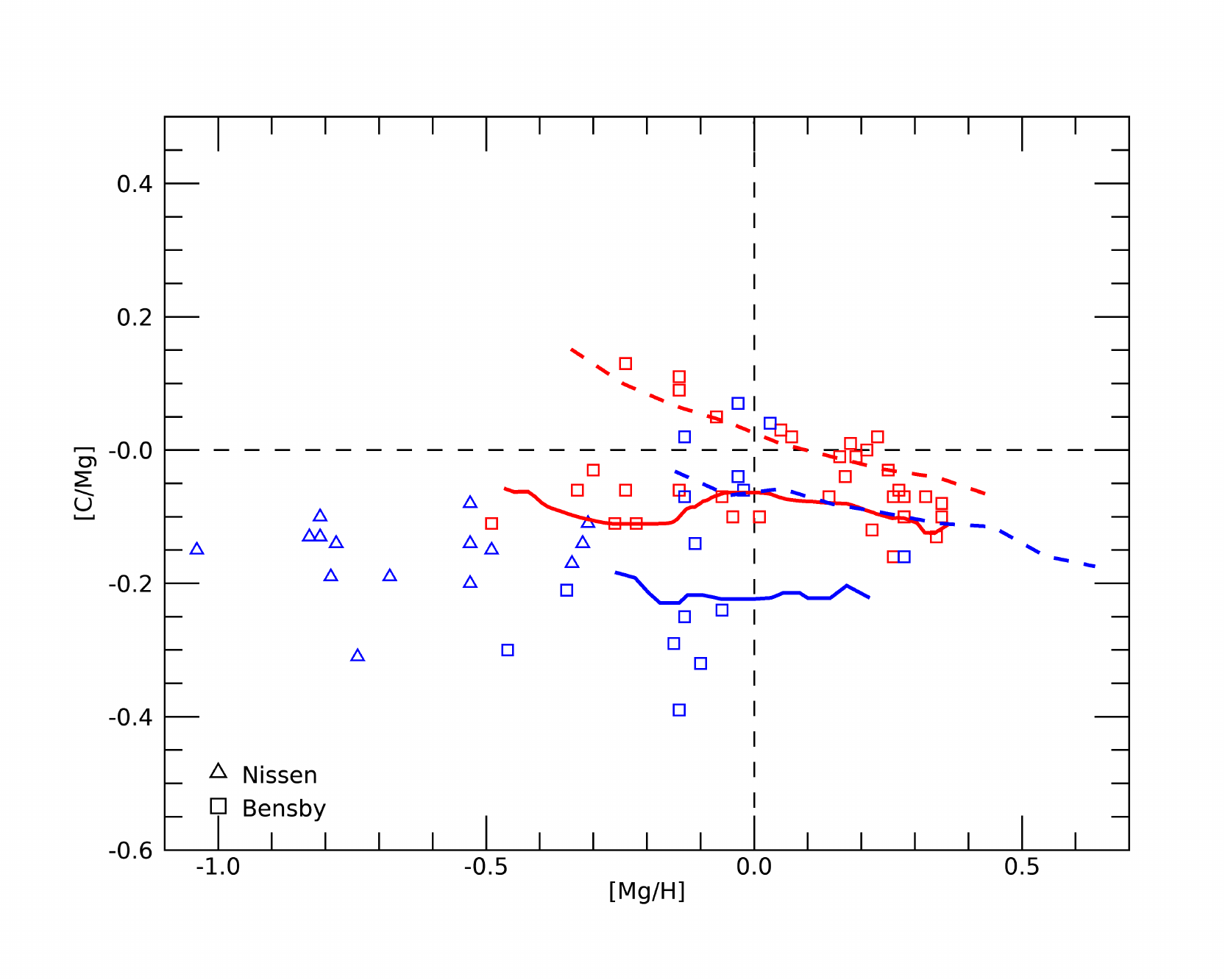}
     \caption{ Carbon-to-iron (\emph{left panel}) and carbon-to-magnesium 
       (\emph{right panel}) abundance ratios as functions of [Fe/H] and [Mg/H], 
       respectively. The red and blue symbols represent abundance estimates 
       from targeted observations of thin- and thick-disc stars, respectively 
       (triangles: \citealt{2010A&A...511L..10N,2014A&A...568A..25N}; squares: 
       \citealt{2006MNRAS.367.1181B}, for C; \citealt{2014A&A...562A..71B}, for 
       Mg; circles: \citealt{2019A&A...630A.104A}). The red and blue lines 
       represent trends of abundance ratios of thin- and thick-disc stars, 
       respectively, derived from: (i) nearly 1\,300 thin-disc and 100 
       thick-disc dwarf stars from GES iDR5, selected on the bases of chemical 
       criteria \citep[continuous lines;][]{2020ApJ...888...55F}; (ii) more 
       than 12\,000 stars from GALAH DR2, divided into high-Ia and low-Ia 
       sequences \citep[dashed lines;][]{2019ApJ...886...84G}. No zero-point 
       offsets are applied to the data.}
     \label{fig:figComparison}
   \end{figure*}
%-----------------------------------------------------------------

%-------------------------------------- Two column figure
   \begin{figure*}[ht]
     \centering
     \includegraphics[width=0.48\textwidth]{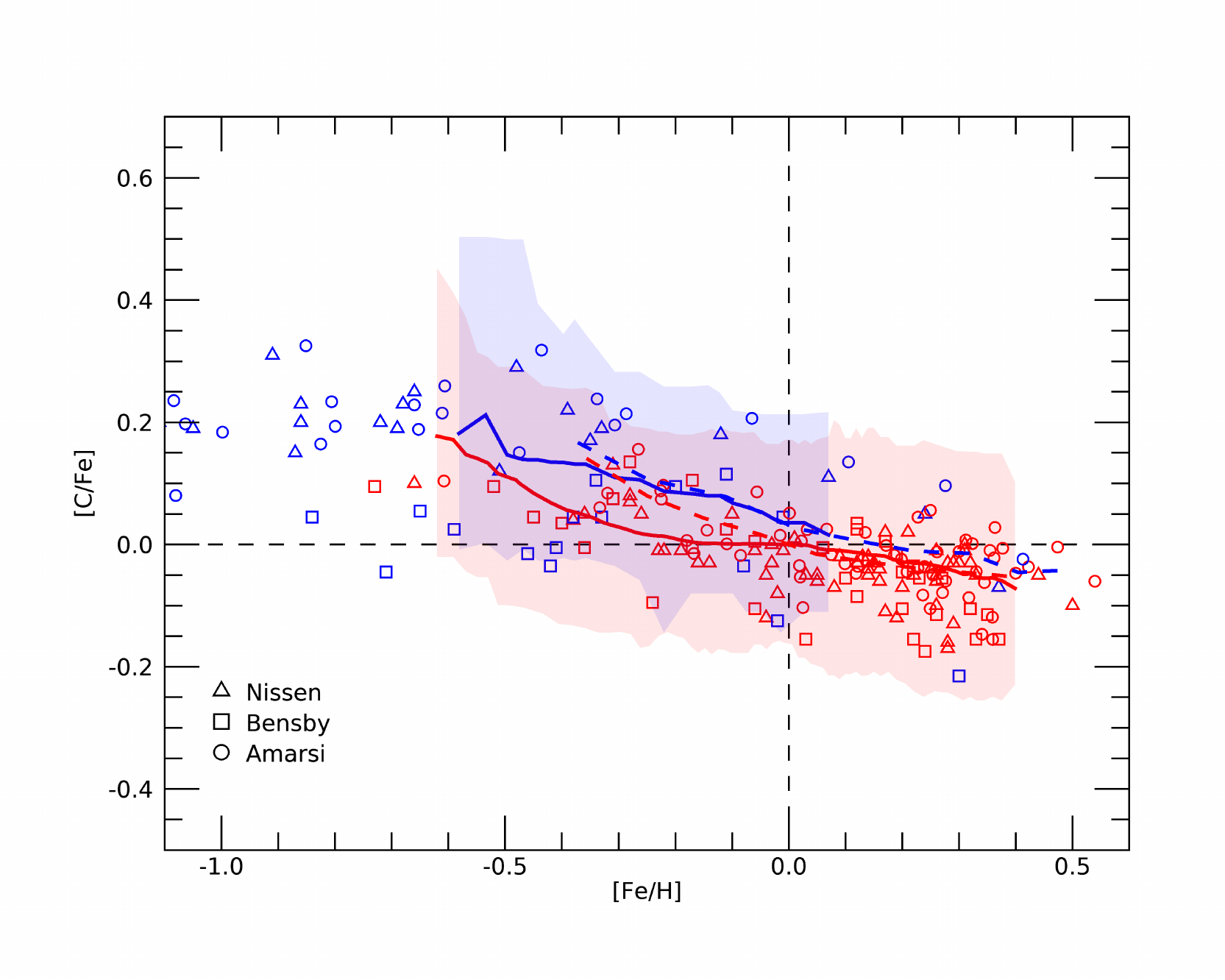}
     \includegraphics[width=0.48\textwidth]{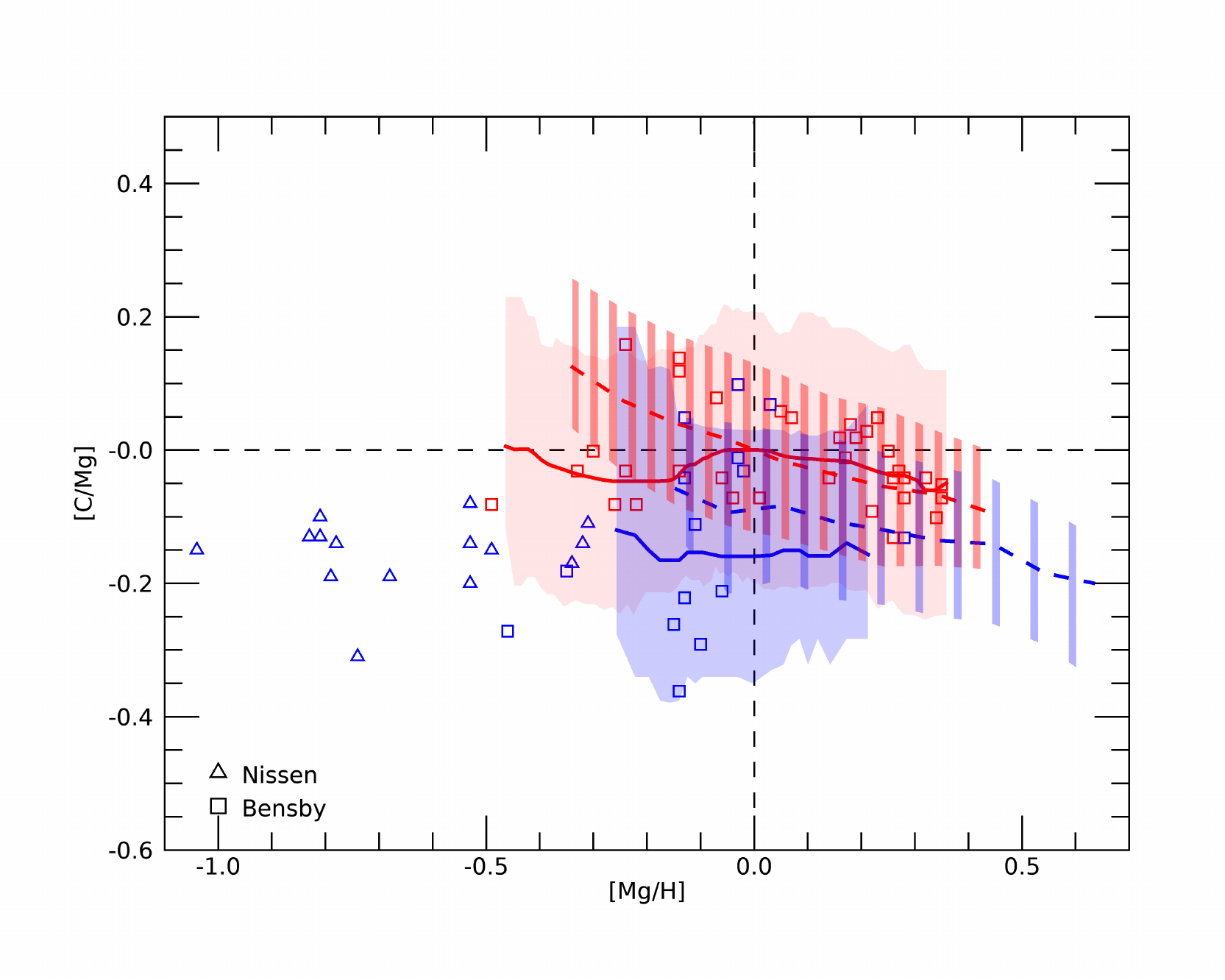}
     \caption{ Same as Fig.~\ref{fig:figComparison}, but with zero-point 
       offsets (see text) applied to the data. Pale red and blue areas show the 
       spread in thin- and thick-disc GES data within boundaries corresponding 
       to the 10th and 90th percentiles. In the right panel, dashed red and 
       blue areas show the analogous spread in the GALAH thin- and thick-disc 
       data, respectively.}
     \label{fig:figScatter}
   \end{figure*}
%-----------------------------------------------------------------

   \subsection{High-resolution spectroscopy of solar vicinity stars}
   \label{sec:hr}

   For the purpose of comparison with our model results, we consider the 
   carbon, oxygen and iron abundances homogeneously derived by 
   \citet{2019A&A...630A.104A} for a sample of 187 F and G dwarfs belonging to 
   the Galactic disc and halo components. The fiducial abundance set is based 
   on a 3D non-LTE line formation analysis, which leads to a reduced scatter 
   and trends at low metallicities that deviate markedly from those obtained in 
   the classical 1D LTE approximation \citep[see][their Figs.~11 to 
     13]{2019A&A...630A.104A}. Consistently with previous work by 
   \citet{2014A&A...568A..25N}, it is found that thick-disc stars have [C/Fe] 
   and [O/Fe] ratios higher than thin-disc stars over the full metallicity 
   range spanned by the observations, with the possible exception of the 
   highest metallicities; in the [C/O] versus [O/H] plane, instead, thin-disc 
   stars display the highest ratios \citep[see, however,][for different 
     results]{2006MNRAS.367.1181B}.

   When high-resolution ($R \ge$~40\,000), high signal-to-noise ($S/N \ge$~150) 
   spectra are available, as it is the case of the studies mentioned above, 
   very small random errors can be associated to single abundance measurements. 
   A significant increase in the sample size (by one or more orders of 
   magnitude) further allows for the definition of tight average trends. 

   \citet{2019ApJ...886...84G} have applied a median trend analysis to more 
   than 70\,000 stars from GALAH DR2; among these, $\sim$12\,000 
   high-metallicity ([Fe/H]~$> -0.4$~dex), hot 
   (5500~$< T_{\rm{eff}}/{\rm{K}} <$~6500) stars have C detections 
   \citep[see][]{2019A&A...624A..19B}. The sample is separated in high-$\alpha$ 
   and low-$\alpha$ stars; these are renamed low-Ia and high-Ia stars, 
   respectively, following common wisdom that a lower/higher level of Fe 
   enrichment from Type Ia supernovae (SNeIa) has to be expected in the 
   thick/thin disc because of its faster/slower evolution \citep[time-delay 
     model,][see also \citealt{1986A&A...154..279M}, their 
     Fig.~3]{1980FCPh....5..287T}. We take advantage of the median low-Ia and 
   high-Ia sequences defined in the [C/X] versus [X/H]\footnote{X~= Fe, O, and 
     Mg.} planes by \citet{2019ApJ...886...84G} to validate our models.

   Moreover, we consider the [C/Fe] versus [Fe/H] and [C/Mg] versus [Mg/H] 
   average behaviours for thick- and thin-disc stars selected on a chemical 
   basis from GES iDR5 by \citet{2020ApJ...888...55F}. These authors have 
   re-analysed spectra taken with UVES in the setup centred at 580~nm 
   ($R \simeq$~47\,000) and derived carbon abundances from atomic lines for 
   2133 FGK stars. The final abundance trends are not corrected for 3D non-LTE 
   effects since, at a first approximation, these prove not to affect the 
   results significantly in the metallicity range spanned by GES observations 
   \citep[see][their Sect.~2.2.2]{2020ApJ...888...55F}.

   In Fig.~\ref{fig:figComparison} we plot for thin- and thick-disc stars the 
   [C/Fe] (left panel) and [C/Mg] (right panel) abundance ratios versus [Fe/H] 
   and [Mg/H], respectively, using abundance estimates from  targeted 
   observations by several authors (triangles: \citealt{2010A&A...511L..10N, 
    2014A&A...568A..25N}; squares: \citealt{2006MNRAS.367.1181B, 
    2014A&A...562A..71B}; circles: 3D non-LTE abundance estimates by 
   \citealt{2019A&A...630A.104A}); the corresponding trends from GES iDR5 
   (continuous lines: \citealt{2020ApJ...888...55F}) and from GALAH DR2 (dashed 
   lines: \citealt{2019ApJ...886...84G}) are also shown. Red and blue colours 
   are adopted to identify thin- and thick-disc stars, respectively. From 
   Fig.~\ref{fig:figComparison} we can clearly observe significant differences 
   between the  trends from the two surveys including systematic offsets. The 
   GES trends are lower than the GALAH  ones and neither of them agree with the 
   loci of the individual points, even if they are characterised by a large 
   spread. Actually, the comparison between abundances derived by different 
   studies is always difficult because several factors (e.g., different 
   resolution and $S/N$, atmospheric parameters uncertainties, different 
   methods of analysis, etc.) may introduce large systematic differences.

   A common approach in literature is to remove at least systematic offsets by 
   introducing zero-point corrections so that thin-disc stars with 
   [Fe/H]~$\simeq$~0 (or [Mg/H]~$\simeq$~0) also have solar chemical 
   composition (i.e., [X/Fe] or [X/Mg]~$\simeq$~0). As an example, 
   \citet{2019ApJ...886...84G} apply zero-point offsets to their GALAH DR2 data 
   points and trends so that stars with [Fe/Mg]~=~[Mg/H]~$\simeq$~0 also have 
   [X/Mg]~$\simeq$~0. We adopted a similar approach and computed for each group 
   of data shown in Fig.~\ref{fig:figComparison} zero-point offsets in order to 
   have [C/Fe] and [C/Mg]~$\simeq$~0 for thin-disc stars with [Fe/H] or 
   [Mg/H]~$\simeq$~0, respectively. Figure~\ref{fig:figScatter} shows the same 
   comparison as in Fig.~\ref{fig:figComparison} but after applying these 
   zero-point offsets. From the left panel we can see that the agreement 
   between the offset [C/Fe] GES and GALAH trends is good but for the 
   metal-poor thin-disc stars, where the GALAH trend falls at higher [C/Fe] 
   than the GES one. The latter seems in better agreement with the area covered 
   by individual points than the former, but the large scatter of individual 
   abundance ratios prevents us to draw a clear conclusion.

   The spread in the thin and thick GES [C/Fe] abundance ratios is shown by the 
   pale red and blue areas, respectively, with boundaries corresponding to the 
   10th to 90th percentiles. The quite large vertical extension of these two 
   areas ($\simeq \pm 0.2$~dex) may come from a combination of uncertainties in 
   the derived abundance ratios (due to the relatively low $S/N$ of some of the 
   GES spectra) and of a possible cosmic scatter. On the other hand, the fact 
   that also the individual points, which are affected by much smaller 
   uncertainties, span similar wide areas seems to indicate that the cosmic 
   scatter may be significant for both thin- and thick-disc stars. A similar 
   result was found for [O/Fe] in the Galactic disc by 
   \citet{2016A&A...590A..74B} who found that the square root of the 
   star-to-star cosmic variance in [O/Fe] ratio at a given metallicity is about 
   0.03--0.04~dex in both the thin and thick disc.

   In the right panel of  Fig.~\ref{fig:figScatter} we plot the same quantities 
   as in the left panel but for [C/Mg] versus [Mg/H]. Moreover, we plot in this 
   panel as red and blue vertical shaded regions the spread areas, between the 
   10th and 90th percentiles, covered by GALAH data of thin- and thick-disc 
   stars as provided to us by E. Griffith (private communication). It can be 
   seen that the GALAH spread areas are comparable in size with the GES ones 
   taking also into account the different number of stars in the two samples. 
   Also in this diagram the scatter of the individual points is similar to 
   those of the surveys, thus supporting the presence of a cosmic scatter in 
   [C/Mg] as in [C/Fe].
 
   Even if by applying zero-point offsets the discrepancies between GES and 
   GALAH trends showed in the right panel of Fig.~\ref{fig:figComparison} are 
   reduced, significant differences still remain in Fig.~\ref{fig:figScatter}. 
   In particular, [C/Mg] GES trends are flatter than GALAH ones and in better 
   agreement with the loci of the individual points, especially for the 
   thick-disc stars. It may be worthwhile pointing out that flat trends of 
   [C/Mg] are expected if C and Mg are mainly produced by the same progenitors 
   (see discussion in Sect.~\ref{sec:disc}).

   \subsection{Microlensed G dwarfs in the Galactic bulge}

   The targets of the high-resolution studies reviewed in the previous section 
   are all main-sequence or subgiant stars. The original carbon content of 
   their atmospheres is not altered by internal evolutionary processes and as 
   such, they provide a fossil record of the C enrichment histories of the 
   different Galactic components they belong to. However, they are faint and 
   probe mostly the immediate surroundings of the Sun.

   In principle, high-resolution spectra can be obtained also for dwarf stars 
   in the Galactic bulge, if they are optically magnified during gravitational 
   microlensing events. The stars can then be analysed the same way as control 
   sample analogs in the solar neighbourhood, thus enabling a fully 
   differential comparison between different Galactic populations 
   \citep[e.g.,][]{2010A&A...512A..41B}. In practice, to the best of our 
   knowledge carbon abundances have been derived only for three microlensed 
   dwarfs in the bulge\footnote{T.~Bensby and collaborators have collected a 
     statistically significant sample of 91 microlensed dwarf and subgiant 
     stars in the bulge \citep[][and references therein]{2017A&A...605A..89B}. 
     They provide stellar ages and abundances for several elements, but do not 
     encompass carbon: that's a bummer!}, namely, MOA--2006--BLG--099S 
   \citep{2008ApJ...685..508J}, MOA--2008--BLG--310S and MOA--2008--BLG--311S 
   \citep{2009ApJ...699...66C}.

   \subsection{Early-type galaxies}

   Along with the Galactic bulge, massive elliptical galaxies are the stellar 
   systems where one expects to see the cleanest signs of nucleosynthesis at 
   high metallicity.

   From an analysis of unresolved stellar populations in nearly 4\,000 
   early-type galaxies from the Sloan Digital Sky Survey 
   \citep[SDSS,][]{2000AJ....120.1579Y}, \citet{2012MNRAS.421.1908J} find that 
   [C/Mg] is about solar for the most massive systems, while [C/O] is slightly 
   higher, namely [C/O]~$\sim$~0.05 \citep[see also][]{2014ApJ...780...33C}. It 
   is suggested that similar values of the abundance ratios imply a lower limit 
   for the formation timescale of massive ellipticals of about 0.4~Gyr, which 
   is the lifetime of a 3~M$_\odot$ star (the underlying hypothesis is that in 
   order to explain the observed ratios intermediate-mass stars must contribute 
   their newly-produced carbon). In Sect.~\ref{sec:disc} we revise this 
   assertion in the light of updated stellar nucleosynthesis and chemical 
   evolution calculations.

   \section{Chemical evolution models}
   \label{sec:models}

%-----------------------------------------------------------------
   \begin{table*}
     \caption{Input parameters for different models.}
     \label{tab:tab1}
     \centering
     \begin{tabular}{l c c c c}
       \hline\hline
       Model & e-folding time in infall law & Delay time for infall & Star formation efficiency & gwIMF \\
    & ($\tau_1$, $\tau_2$ or $\tau$ [Gyr]) & ($t_{\mathrm{max}}$ [Gyr]) & ($\nu$ or $\tilde{\nu}$ [Gyr$^{-1}$]) \\
       \hline
       Two-infall, thick disc (classical) &  1.0 & --  &  2.0 & Canonical\tablefootmark{a} \\
       Two-infall, thin disc (classical)  &  7.0 & 1.0 &  1.0 & Canonical\tablefootmark{a} \\
       Two-infall, thick disc (revised)   &  0.1 & --  &  1.3 & Canonical\tablefootmark{a} \\
       Two-infall, thin disc (revised)    &  8.0 & 4.3 &  1.3 & Canonical\tablefootmark{a} \\
       Parallel, thick disc               &  0.5 & --  &  2.0 & Canonical\tablefootmark{a} \\
       Parallel, thin disc                &  7.0 & --  &  1.0 & Canonical\tablefootmark{a} \\
       Galactic bulge                     &  0.1 & --  & 25.0 & Salpeter\tablefootmark{b}  \\
       Prototype elliptical galaxy        & 0.05 & --  &  1.8 & Top-heavy\tablefootmark{c}  \\
       \hline
     \end{tabular}
     \tablefoot{
       \tablefoottext{a}{\citet{2002ASPC..285...86K} IMF with $x =$~1.7 in the 
         high-mass domain ($x =$~1.35 for Salpeter IMF).}
       \tablefoottext{b}{Extrapolated \citet{1955ApJ...121..161S} slope over 
         the whole mass range.}
       \tablefoottext{c}{Slope flatter than Salpeter's, $x =$~1.1, in the 
         high-mass range 
         \citep[see][]{2018Natur.558..260Z,2019MNRAS.490.2838R}.}
     }
   \end{table*}
%-----------------------------------------------------------------

   In this work, we adopt the following chemical evolution models:
   \begin{enumerate}
   \item The two-infall chemical evolution model for the Milky Way halo and 
     discs, originally developed by \citet{1997ApJ...477..765C,
       2001ApJ...554.1044C}. This model divides the Galactic disc in several 
     concentric annuli 2~kpc wide that evolve independently, without exchanges 
     of matter between them \citep[but see][]{2015ApJ...802..129S}. The inner 
     halo and thick disc are assumed to form fast, in less than 1~Gyr, out of a 
     first episode of accretion of virgin gas. During these earlier 
     evolutionary stages, the star formation proceeds very efficiently and 
     turns a large fraction of the available gas into stars. Eventually, a 
     critical gas density threshold is reached below which the star formation 
     halts. Later on, a second infall episode, delayed 1~Gyr with respect to 
     the previous one, starts replenishing the disc with fresh gas and star 
     formation is reignited. The thin disc thence forms on longer timescales 
     that are functions of the Galactocentric distance 
     \citep{1989MNRAS.239..885M,2000ApJ...539..235R}. This scheme has been 
     recently revised by \citet{2019A&A...623A..60S} who have taken into 
     account the constraints on stellar ages provided by asteroseismology and 
     fixed the delay time for the second infall to 4.3~Gyr, i.e., much longer 
     than assumed in the original formulation of the model. Here we show 
     results obtained for the solar neighbourhood in either case.
   \item The parallel model proposed by \citet{2017MNRAS.472.3637G} that 
     envisages distinct formation sequences for the thick and thin discs 
     \citep[see also][]{2019MNRAS.489.3539G,2020MNRAS.492.2828G}. In the 
     parallel model, the two discs form independently on different timescales 
     out of two separate infall episodes. In this work we consider only the 
     formation of the local discs (i.e., within 1~kpc from the Sun), but the 
     model has been extended to other Galactocentric distances 
     \citep{2018MNRAS.481.2570G}.
   \item The reference model for the Galactic bulge by 
     \citet{2019MNRAS.487.5363M}, which assumes that the majority of the stars 
     in the so-called `classical bulge' are old and form quickly in a very 
     efficient starburst. These assumptions, jointly to the adoption of a gwIMF 
     flatter than the one derived for the solar vicinity, guarantee that the 
     stellar metallicity and age distributions, as well as the [Mg/Fe] ratios, 
     are well reproduced by the model \citep[][see also 
       \citealt{1990ApJ...365..539M,1999A&A...341..458M}]{2019MNRAS.487.5363M}.
   \item The model for the prototype massive elliptical galaxy described in 
     \citet{2017MNRAS.470..401R,2019MNRAS.490.2838R}. According to these 
     authors, the most massive early-type galaxies form stars intensively while 
     hidden in heavy dust curtains at high redshifts, where they show up as 
     submillimeter-bright galaxies. After reaching a stellar mass of about 
     $2 \times 10^{11}$~M$_\odot$, their residual gas is swept away by the 
     cleaning action of large-scale outflows triggered by SN explosions and AGN 
     activity. The galaxies evolve passively since then. Similarly to what is 
     done for the Galactic bulge, a top-heavy gwIMF is assumed, which 
     guarantees a good fit to the CNO isotopic ratios measured in submillimeter 
     galaxies \citep{2017MNRAS.470..401R,2018Natur.558..260Z}.
   \end{enumerate}

   \subsection{Basic assumptions}
   \label{sec:bas}

   The chemical evolution models adopted in this study track the evolution of 
   the chemical composition of the interstellar medium (ISM) in different 
   galaxies and/or Galactic components. They deal with several elements, from 
   the lightest ones emerging from Big Bang nucleosynthesis 
   \citep{2003MNRAS.346..295R} to the heaviest ones synthesised by uncommon 
   astrophysical sites, like magneto-rotational driven SNe, and/or through rare 
   events, as compact binary mergers \citep{2014MNRAS.438.2177M,
     2015A&A...577A.139C,2020MNRAS.492.2828G}.

   Cold gas of primordial chemical composition is accreted at an exponentially 
   decreasing rate. For the two-infall model,
   \begin{equation}
     \frac{{\rm d}\mathscr{M}_{\rm inf}(t)}{{\rm d}t} = c_1 \, {\rm e}^{-t/\tau_1} 
                                       + c_2 \, {\rm e}^{-(t-t_{\mathrm{max}})/\tau_2},
   \end{equation}
   where $\mathscr{M}_{\rm inf}(t)$ is the mass accreted at time $t$, $\tau_1$ 
   and $\tau_2$ are the e-folding times of the first and second infall 
   episodes, respectively, and $t_{\mathrm{max}}$ indicates the delay of the 
   beginning of the second infall. The coefficients $c_1$ and $c_2$ are 
   obtained by imposing a fit to the currently observed surface mass densities 
   of the thick- and thin-disc components in the solar neighbourhood 
   (obviously, $c_2$ is set to zero if $t < t_{\mathrm{max}}$). For the other 
   models the formula gets simpler,
   \begin{equation}
     \frac{{\rm d}\mathscr{M}_{\rm inf}(t)}{{\rm d}t} = c \, {\rm e}^{-t/\tau},
   \end{equation}
   where the different quantities have the usual meaning.

   In all models, the star formation rate is implemented according to the 
   Kennicutt-Schmidt law \citep{1959ApJ...129..243S,1998ApJ...498..541K}. In 
   the models for the Galactic bulge, halo and discs it reads
   \begin{equation}
     \psi(r, t) = \nu \, \left[ \frac{\sigma_{\mathrm{tot}}(r, t) \, 
         \sigma_{\mathrm{gas}}(r, t)}{\sigma_{\mathrm{tot}}(r_\odot, t)^2} 
       \right]^{(k-1)} \sigma_{\mathrm{gas}}^k(r, t),
   \end{equation}
   where $\sigma_{\mathrm{gas}} (r, t)$ is the surface gas density at a given 
   radius and time, $\sigma_{\mathrm{tot}} (r, t)$ is the total surface mass 
   density at a given radius and time, $\sigma_{\mathrm{tot}}(r_\odot, t)$ is the 
   total surface mass density at the solar position, and $k =$~1.5. In the 
   one-zone model for ellipticals
   \begin{equation}
     \psi(t) = \tilde{\nu} \, \mathscr{M}_{\mathrm{gas}}^k(t),
   \end{equation}
   where $\mathscr{M}_{\mathrm{gas}} (t)$ is the cold gas mass at time $t$ and 
   $k =$~1. Either way, the proportionality constants ($\nu$, $\tilde{\nu}$) 
   are free parameters and vary from model to model.

   As in previous works, the gwIMF is the canonical one \citep[][with $x =$ 1.7 
     in the high-mass domain]{2002ASPC..285...86K} for the solar neighbourhood 
   models, while flatter gwIMFs are assumed in the models for spheroids.

   Finally, galactic winds are considered only in the model for early-type 
   galaxies.

   The values of the main model parameters adopted in this work are summarized 
   in Table~\ref{tab:tab1}.

   \subsection{Nucleosynthesis prescriptions}
   \label{sec:nuc}

   The models presented in this paper account in detail for the finite stellar 
   lifetimes, namely, the instantaneous recycling approximation is relaxed. 
   However, they assume instantaneous, homogeneous mixing of the stellar ejecta 
   within the ISM. As a consequence, these models can only predict average 
   trends, while nothing can be said about the spread in the data.

   As for the two-infall model, the adopted nucleosynthesis prescriptions are 
   the same as model MWG-11 of \citet{2019MNRAS.490.2838R}, namely, we adopt 
   the yields by \citet{2013MNRAS.431.3642V} for single low- and 
   intermediate-mass stars and the yields by \citet{2018ApJS..237...13L} for 
   high-mass stars. In particular, in model MWG-11 the initial rotational 
   velocities of massive stars are allowed to vary from the maximum value 
   considered by \citet{2018ApJS..237...13L}, i.e., 300~km~s$^{-1}$, for 
   [Fe/H]~$< -$1 to zero at higher metallicities. Furthermore, the mass range 
   for full collapse to black holes is set to 30--100~M$_\odot$ for 
   [Fe/H]~$< -$1, but is reduced to 60--100~M$_\odot$ for [Fe/H]~$\ge -$1. 
   This leads to a good agreement between model predictions and observations 
   of CNO element abundances and isotopic ratios in the solar vicinity, as 
   well as across the whole Galactic disc, in the framework of our model 
   \citep[see][]{2019MNRAS.490.2838R}.

   All other models are run three times, one for each choice of nucleosynthetic 
   yields corresponding to the assumptions of models~MWG-05, MWG-06 and MWG-07 
   of \citet{2019MNRAS.490.2838R}. In brief, we adopt the yields from 
   \citet{2013MNRAS.431.3642V} for (non-rotating) low- and intermediate-mass 
   stars and the yields from \citet[][their recommended 
     set~R]{2018ApJS..237...13L} with initial rotational velocities of 0 
   (models labeled `0 km/s' in Figs.~\ref{fig:fig2} to \ref{fig:fig6}), 150 
   (models labeled `150 km/s' in Figs.~\ref{fig:fig2} to \ref{fig:fig6}), or 
   300~km~s$^{-1}$ (models labeled `300 km/s' in Figs.~\ref{fig:fig2} to 
   \ref{fig:fig6}) for stars dying as core-collapse SNe. The interested reader 
   is addressed to \citet[][and references therein]{2019MNRAS.490.2838R} for 
   further details on the adopted stellar yield sets.

   The nucleosynthetic outcome of SNeIa is also included in the models. In 
   particular, we adopt the yields by \citet{1999ApJS..125..439I} and the 
   single-degenerate scenario for the progenitors following 
   \citet{1986A&A...154..279M} and \citet{2001ApJ...558..351M}. The 
   contribution to C, O and Mg synthesis from SNeIa is negligible, while they 
   produce huge amounts of Fe.

   \section{Results}
   \label{sec:results}

%-------------------------------------- Two column figure
   \begin{figure*}
     \centering
     \includegraphics[width=0.48\textwidth]{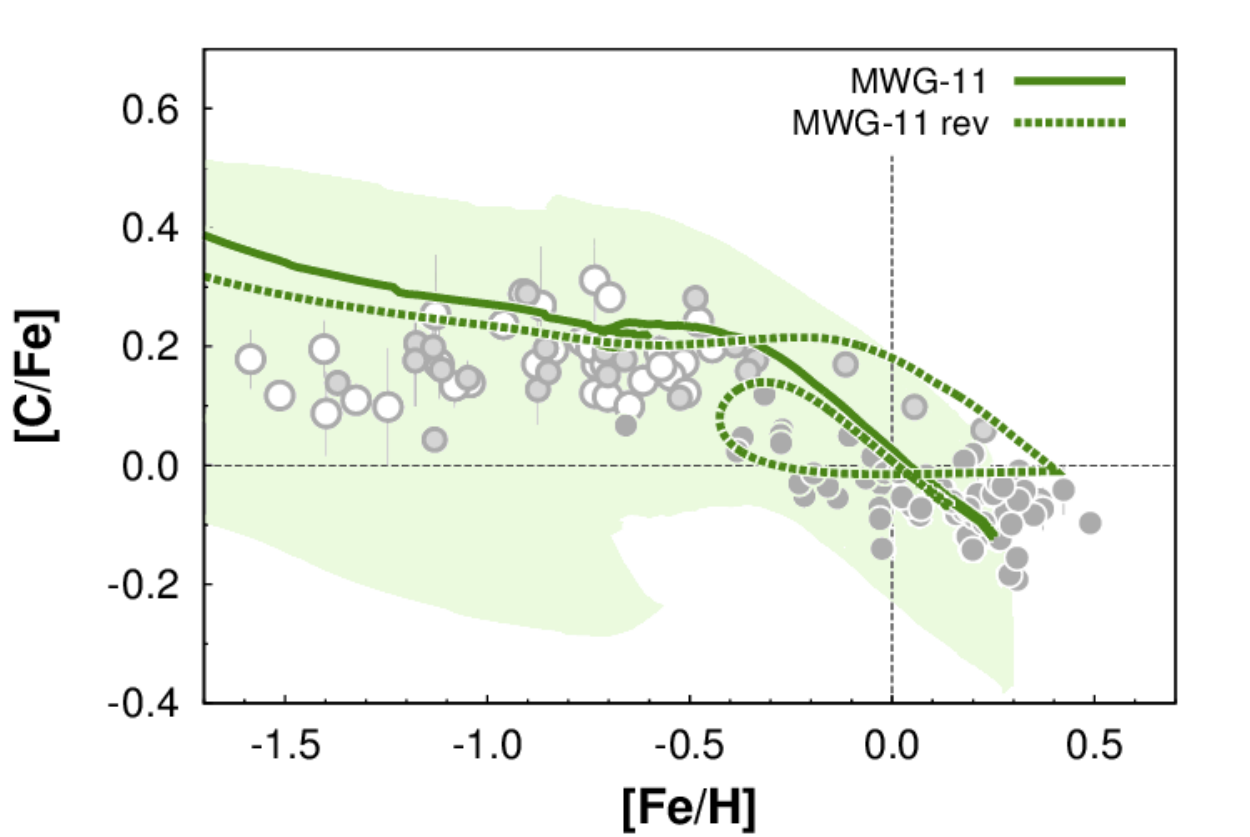}
     \includegraphics[width=0.48\textwidth]{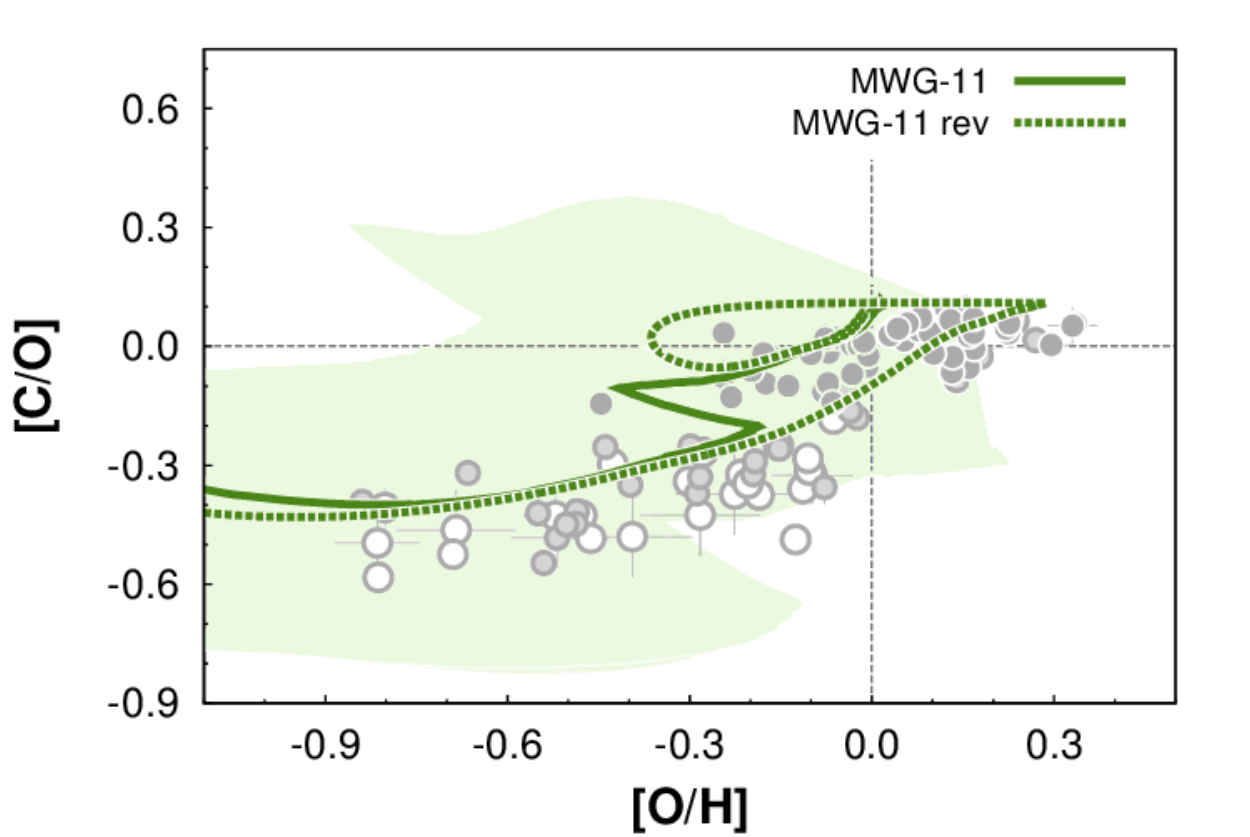}
     \caption{ Carbon-to-iron (\emph{left panel}) and carbon-to-oxygen 
       (\emph{right panel}) abundance ratios as functions of [Fe/H] and [O/H], 
       respectively, in the solar vicinity. The solid lines show the 
       predictions of model MWG-11 by \citet{2019MNRAS.490.2838R}; theoretical 
       uncertainties arising from stellar nucleosynthesis are highlighted by 
       pale green areas. The dashed lines show the predictions of model MWG-11 
       revised following \citet[][see text and 
         Table~\ref{tab:tab1}]{2019A&A...623A..60S}. The symbols represent 3D 
       non-LTE abundance estimates from high-resolution spectra of thin-disc, 
       thick-disc and high-$\alpha$ halo stars \cite[dark grey filled circles, 
         light grey filled circles and empty circles, 
         respectively;][]{2019A&A...630A.104A}; in most cases, the 
       observational error lies within the symbol size. All abundance ratios 
       are normalised to the solar photospheric composition by 
       \citet{2009ARA&A..47..481A}.}
   \label{fig:fig1}
   \end{figure*}
%-----------------------------------------------------------------

   \subsection{The solar vicinity}
   \label{sec:sv}

   \subsubsection{Two-infall model predictions}

   In Fig.~\ref{fig:fig1}, left panel, we show [C/Fe] abundance ratios against 
   [Fe/H], on the right-hand side, we display the [C/O] versus [O/H] diagram. 
   The circles refer to abundance determinations from high-resolution spectra 
   of single dwarfs in the solar neighbourhood, after corrections for 3D and 
   non-LTE effects \citep{2019A&A...630A.104A}. The solid lines are the 
   predictions of the best fitting model by \citet{2019MNRAS.490.2838R}, i.e., 
   their model MWG-11. Theoretical uncertainties associated with changes in the 
   adopted stellar yields are highlighted (shaded areas); though they are 
   large, the direct comparison with a good data set provides a sensible way to 
   discriminate among various nucleosynthesis prescriptions. The dashed lines 
   refer to the predictions of model MWG-11 after introducing the improvements 
   to the two-infall evolutionary scheme recently suggested by 
   \citet{2019A&A...623A..60S}.

   In the metallicity range $-1.4 < \mathrm{[Fe/H]} < -0.4$ 
   ($-0.9 < \mathrm{[O/H]} < 0.0$), objects classified as high-$\alpha$ 
   halo stars\footnote{These are possibly stars formed \emph{in situ}, as 
     opposed to low-$\alpha$ halo stars that have likely been accreted from 
     disrupting dwarf satellites \citep[see][and references 
       therein]{2014A&A...568A..25N}.} or thick-disc stars overlap each other. 
   At variance with halo stars, some thick-disc members are found at higher 
   metallicities, up to [Fe/H]~$\sim 0.2$, or [O/H]~$\sim 0.3$ \citep[see 
     also][]{2007ApJ...663L..13B,2011A&A...535L..11A}. In its classical 
   formulation, the two-infall model cannot accommodate thick-disc stars with 
   [Fe/H]~$> -0.6$ (or [O/H]~$> -0.2$). However, when a longer duration is 
   assumed for the inner-halo/thick-disc phase, as suggested by precise stellar 
   ages provided by asteroseismology \citep[][and references 
     therein]{2019A&A...623A..60S}, stars as metal-rich as 
   $\mathrm{[Fe/H]} \simeq 0.4$ (or $\mathrm{[O/H]} \simeq 0.3$) are predicted 
   to form in the thick-disc component, in full agreement with the 
   observations. As shown in Fig.~\ref{fig:fig1}, the theoretical [C/Fe] and 
   [C/O] ratios of these stars agree well (within 0.06~dex) with the observed 
   ratios. The [C/Fe] and [C/O] ratios of the whole sample, though, appear to 
   be slightly overestimated.

   Overall, the revised MWG-11 model reproduces the bunch of local data 
   satisfactorily well, especially if we consider that a sizeable fraction of 
   the most metal-rich stars in the solar neighbourhood (i.e., those with 
   [Fe/H]~$> 0.25$ dex) might actually be not born locally, but migrated from 
   other radii \citep[e.g.,][see also 
     \citealt{2009MNRAS.396..203S}]{2013A&A...558A...9M,2015A&A...580A.126K}, 
   and would, thus, reflect different evolutionary paths. Indeed, their old 
   ages \citep[e.g.,][and references therein]{2017A&A...600A..70A} fully 
   support this hypothesis.

   \subsubsection{Parallel model predictions}

%-------------------------------------- Two column figure
   \begin{figure*}[ht]
     \centering
     \includegraphics[width=0.48\textwidth]{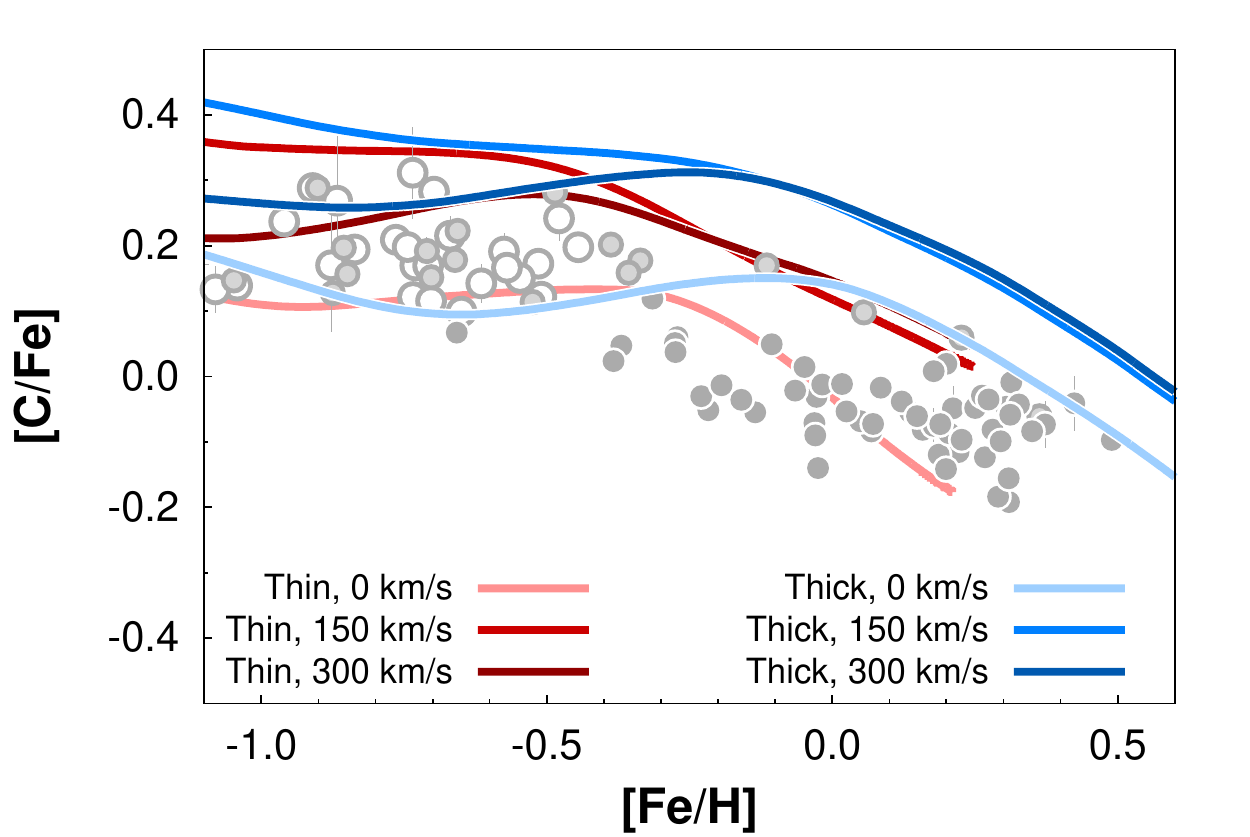}
     \includegraphics[width=0.48\textwidth]{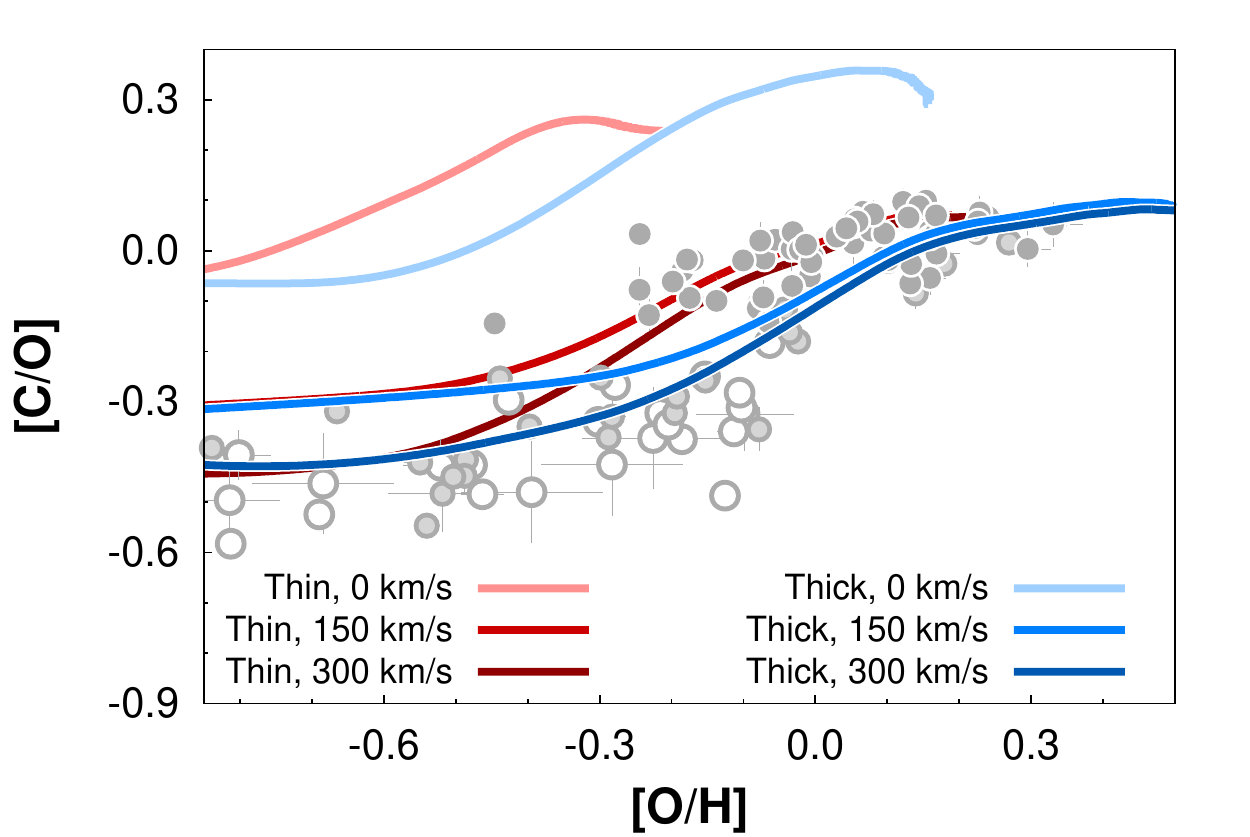}
     \caption{ Similar to the previous figure. Now we show the predictions of 
       the parallel model for thick- and thin-disc formation by 
       \citet{2017MNRAS.472.3637G}, in which we include up-to-date stellar 
       nucleosynthetic yields \citep[see text and][]{2019MNRAS.490.2838R}.}
     \label{fig:fig2}
   \end{figure*}
%-----------------------------------------------------------------

%-------------------------------------- Two column figure
   \begin{figure*}[ht]
     \centering
     \includegraphics[width=0.48\textwidth]{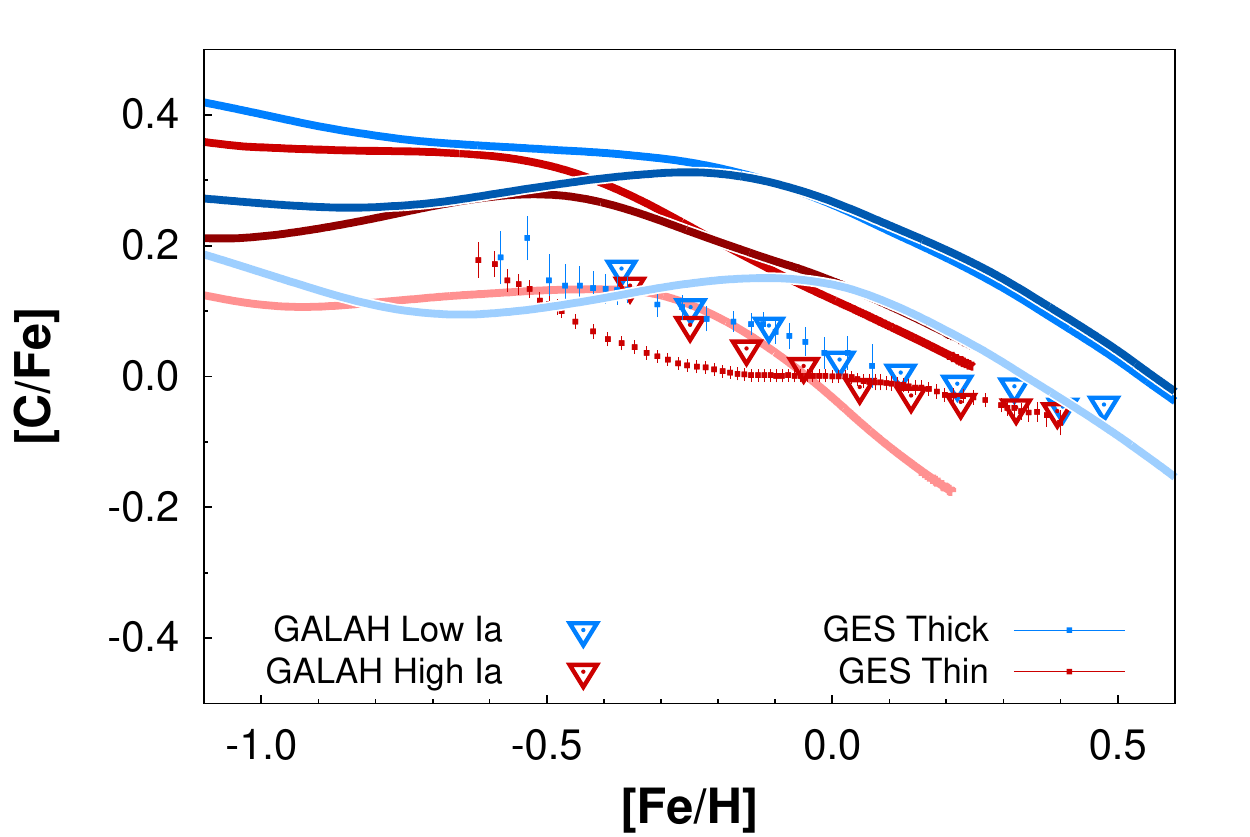}
     \includegraphics[width=0.48\textwidth]{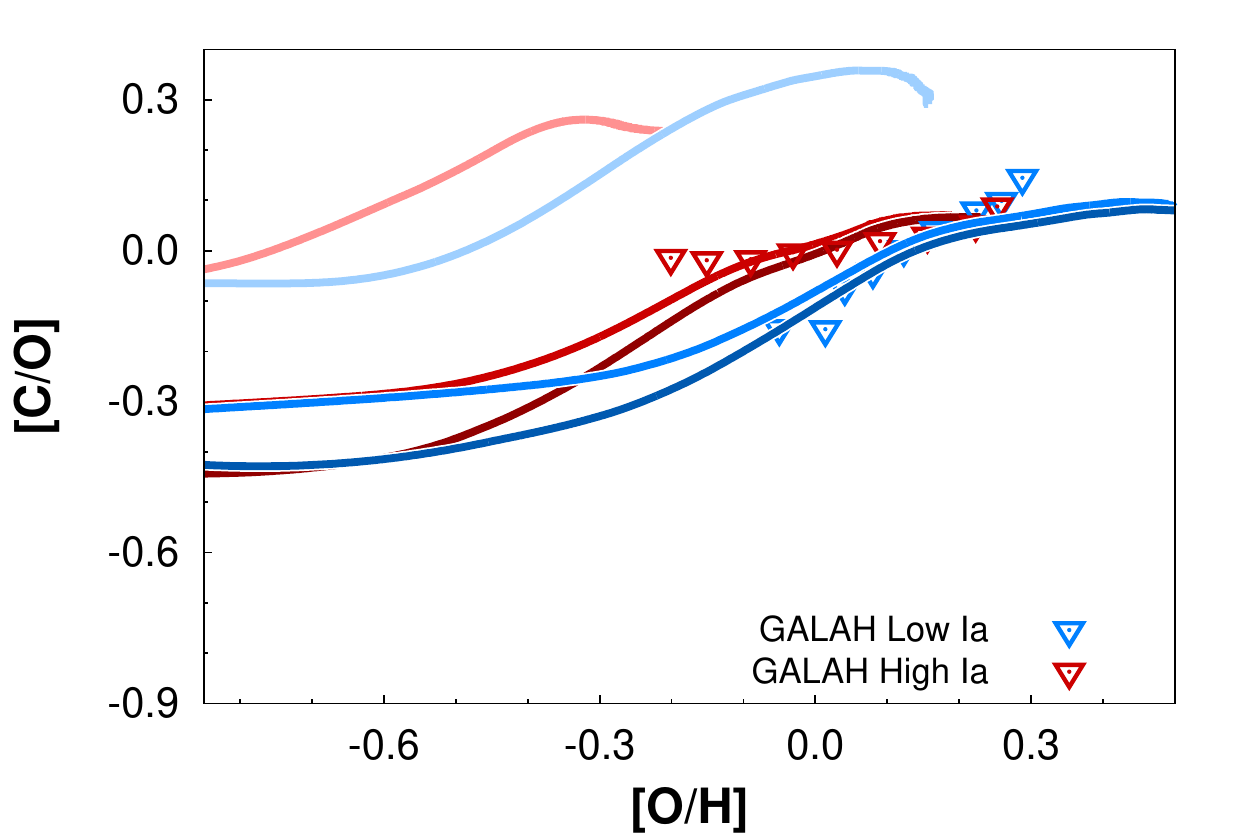}
     \caption{ Same as Fig.~\ref{fig:fig2}, but now the model predictions are 
       compared to the observed trends inferred for: (i) thin- and thick-disc 
       dwarf stars from GES iDR5 \citep[red and blue dots, 
         respectively; ][]{2020ApJ...888...55F}; (ii) high-Ia and low-Ia GALAH 
       DR2 stars \citep[red and blue upside-down triangles, 
         respectively; ][]{2019ApJ...886...84G}. Zero-point shifts are applied 
       to the data (see Sect.~\ref{sec:hr}).}
     \label{fig:fig3}
   \end{figure*}
%-----------------------------------------------------------------

%-------------------------------------- One column figure
   \begin{figure}[ht]
     \centering
     \includegraphics[width=0.48\textwidth]{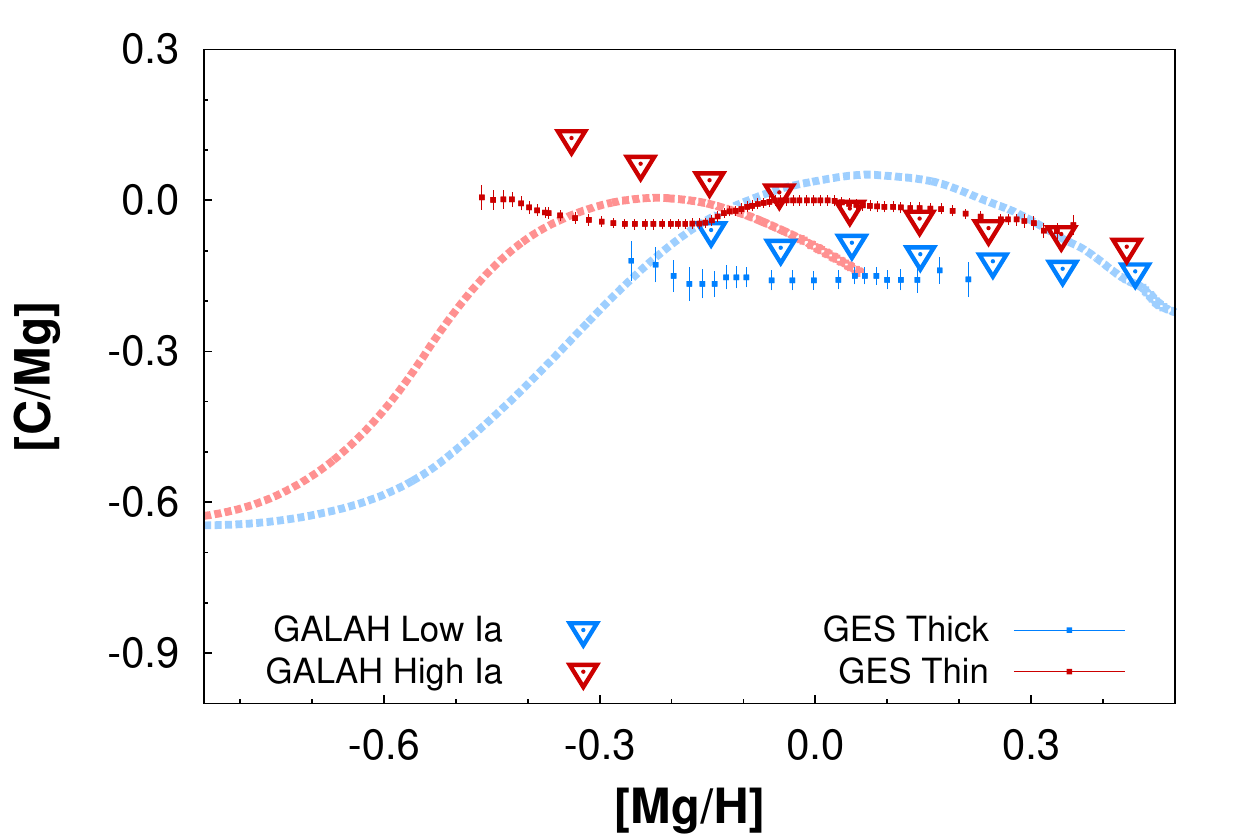}
     \caption{ [C/Mg]--[Mg/H] diagram. Data are from GES iDR5 and GALAH DR2 
       (dots and upside-down triangles, respectively) and include zero-point 
       shifts (see text). Predicted trends for the thick (blue line) and thin 
       (red line) discs are from the parallel model \citep{2017MNRAS.472.3637G} 
       with nucleosynthesis prescriptions as model MWG-01 of 
       \citet{2019MNRAS.490.2838R}.}
     \label{fig:fig4}
   \end{figure}
%-----------------------------------------------------------------

   In the previous section, we discuss the results of model~MWG-11, where an 
   abrupt transition is supposed to occur between a regime that favours the 
   formation of massive fast rotators at low metallicities and a regime that 
   leads mostly to slowly- or non-rotating stars above [Fe/H]~$= -1.0$ 
   \citep{2019MNRAS.490.2838R}. These, or similar, assumptions allow to obtain 
   a satisfactory fit to the abundance distribution of most chemical elements 
   in the Galaxy \citep[][see also 
     \citealt{2018MNRAS.476.3432P}]{2019MNRAS.490.2838R}. However, the 
   dependence of stellar rotation on metallicity is basically unknown, and 
   other factors might act as the prime drivers behind the variations in the 
   angular momenta of the proto-star clouds. Therefore, in the following we 
   discuss separately, for each Galactic component, the results of three models 
   that assume different fixed values for the initial rotational velocity of 
   all massive stars.

   In Fig.~\ref{fig:fig2} we show the predictions of the parallel model by 
   \citet{2017MNRAS.472.3637G} obtained when three different sets of stellar 
   yields, corresponding to models MWG-05, MWG-06 and MWG-07 of 
   \citet{2019MNRAS.490.2838R}, are implemented.

   For stars in the mass range 1--8~M$_\odot$, the yields are from \citet[][and 
     private communication]{2013MNRAS.431.3642V} and are computed for six 
   different values of the initial metallicity of the stars, namely, 
   $Z =$~0.0003, 0.001, 0.004, 0.008, 0.018, and 0.04; notably, this includes a 
   set computed for super-solar metallicity stars. The yields for massive stars 
   are from \citet{2018ApJS..237...13L}. In this case, the modelled stars have 
   initial metallicities of [Fe/H]~= $-3, -2, -1,$ and 0 dex, corresponding to 
   $Z = 3.236 \,\times\, 10^{-5}$, $3.236 \,\times\, 10^{-4}$, 
   $3.236 \,\times\, 10^{-3}$, and $1.345 \,\times\, 10^{-2}$; unfortunately, no 
   models are available at super-solar metallicities. While the stellar models 
   of \citet{2013MNRAS.431.3642V} do not include stellar rotation, 
   \citet{2018ApJS..237...13L} provide yields for non-rotating stars, as well 
   as yields for fast rotators. In Fig.~\ref{fig:fig2}, we indicate with the 
   labels `300 km/s', `150 km/s' and `0 km/s' models in which all massive 
   stars rotate with $\vel_{\rm{rot}} =$ 300 or 150~km~s$^{-1}$, or do not 
   rotate at all, respectively.

   In the parallel model, the formation of the thick and thin discs are 
   completely disentangled. The thick-disc model implementing the yields for 
   non-rotating massive stars (light blue lines in Fig.~\ref{fig:fig2}) agrees 
   very well with the sparse data for metal-rich thick-disc stars in the [C/Fe] 
   versus [Fe/H] plane, it marginally agrees with the points at [Fe/H]~$< -$0.2 
   in the same diagram, but it fails by far to reproduce the [C/O] ratios over 
   the full [O/H] range. When rotation is considered (blue and dark blue lines 
   in Fig.~\ref{fig:fig2}), a good agreement is found between model predictions 
   and observations in the [C/O] versus [O/H] plane, but the agreement with the 
   observed [C/Fe] ratios for [Fe/H]~$> -$0.2 sensibly worsens. The transition 
   from a regime favouring fast rotators to a regime where fast rotators are 
   strongly suppressed when moving from low to high metallicities offers, in 
   principle, an elegant solution to this problem \citep[see 
     Fig.~\ref{fig:fig1} and][]{2019MNRAS.490.2838R}, but in order to test this 
   hypothesis properly, a denser grid of massive star yields, comprising 
   results obtained for intermediate values of the initial rotational 
   velocities, is definitely needed. The models for the thin disc 
   (Fig.~\ref{fig:fig2}, bundles of lines in shades of red) strengthen the 
   above conclusions.

%-------------------------------------- Two column figure
   \begin{figure*}
     \centering
     \includegraphics[width=0.48\textwidth]{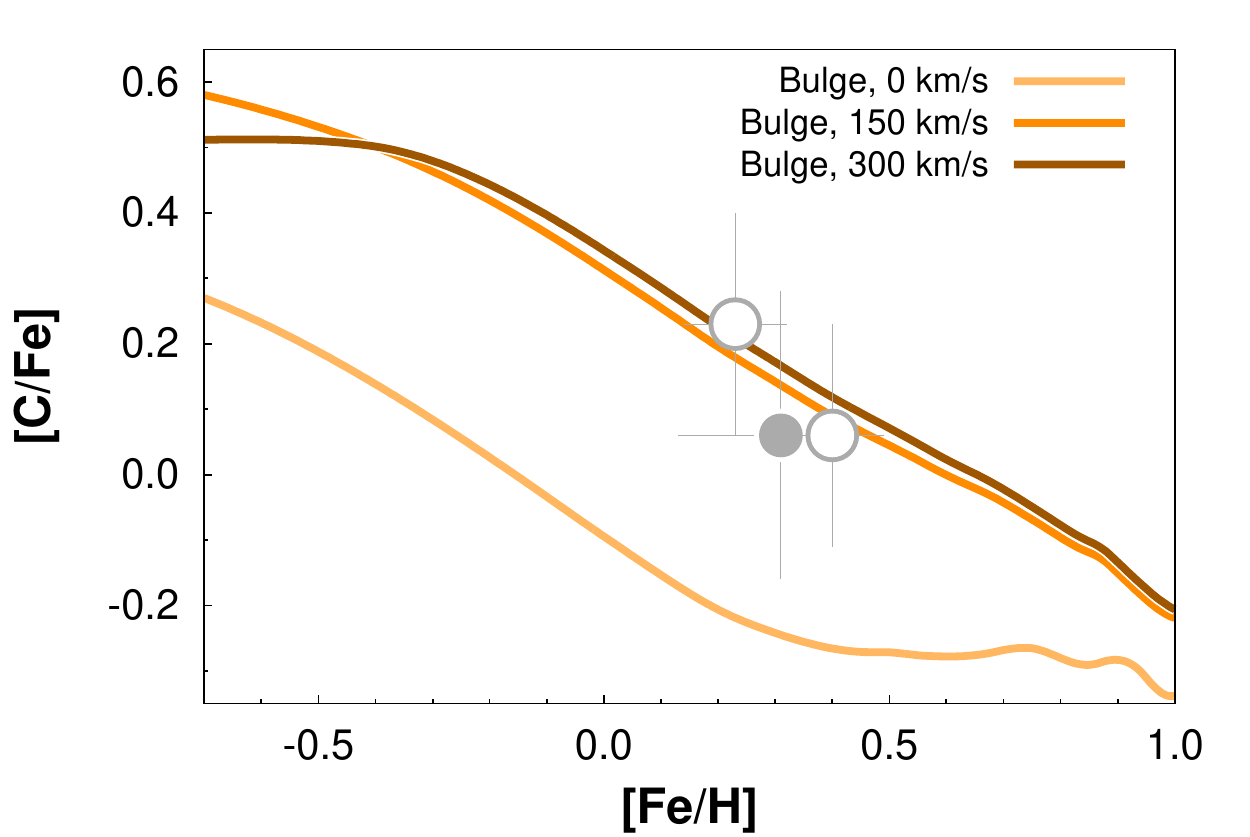}
     \includegraphics[width=0.48\textwidth]{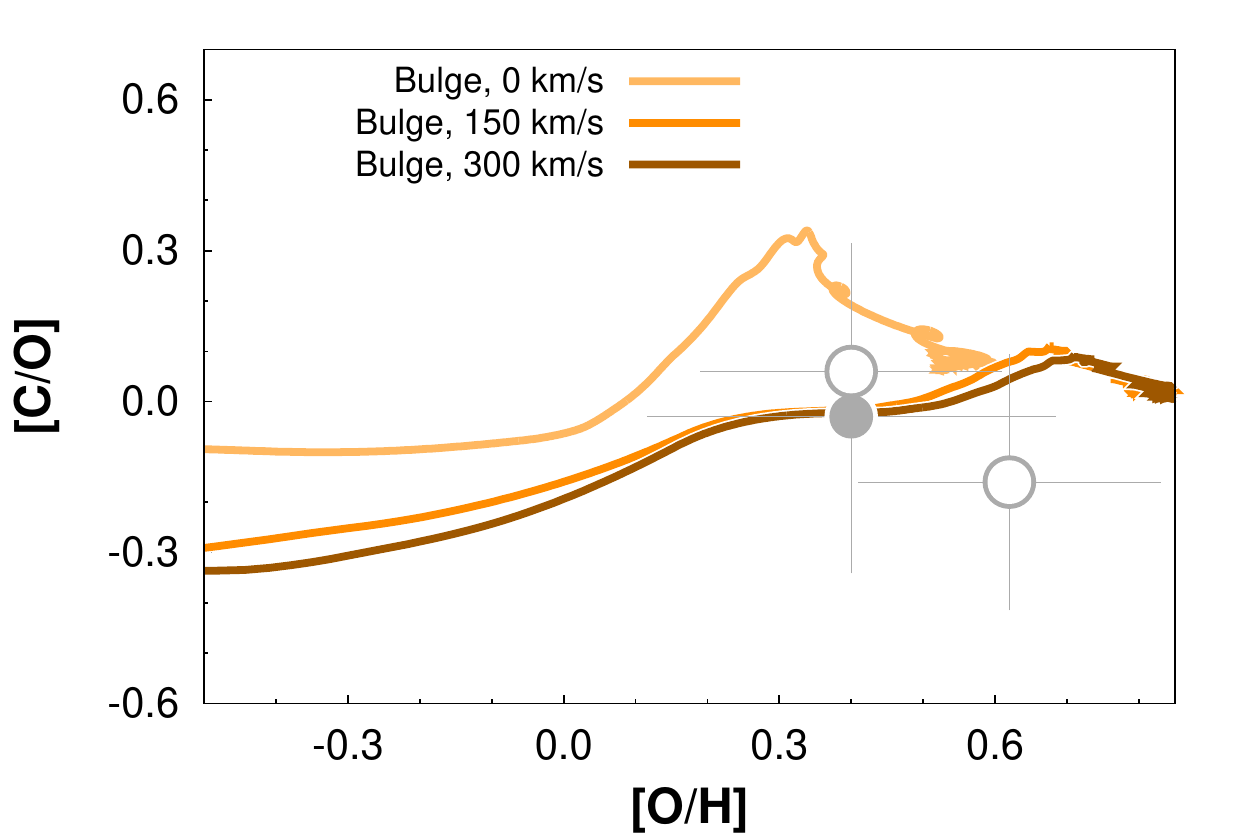}
     \caption{ Carbon-to-iron \emph{(left panel)} and carbon-to-oxygen 
       \emph{(right panel)} abundance ratios as functions of [Fe/H] and [O/H], 
       respectively, in the Galactic bulge. The solid lines show the 
       predictions of the chemical evolution model for the bulge by 
       \citet{2019MNRAS.487.5363M} including the same nucleosynthesis 
       prescriptions as models MWG-05 (brown), MWG-06 (orange), and MWG-07 
       (yellow) of \citet{2019MNRAS.490.2838R}. Data for microlensed bulge 
       dwarfs from \citet[][filled circles]{2008ApJ...685..508J} and 
       \citet[][empty circles]{2009ApJ...699...66C} are also shown. All 
       abundance ratios are normalised to the solar photospheric abundances by 
       \citet{2009ARA&A..47..481A}.}
     \label{fig:fig5}
   \end{figure*}
%-----------------------------------------------------------------

   In principle, nowadays large modern spectroscopic surveys allow us to 
   establish tight abundance trends through the analysis of thousands of stars 
   (see Sect.~\ref{sec:hr}). In Fig.~\ref{fig:fig3}, we compare the same 
   theoretical predictions displayed in Fig.~\ref{fig:fig2} to the low-Ia and 
   high-Ia sequences constructed by \citet{2019ApJ...886...84G} using GALAH DR2 
   data (upside-down triangles), as well as to the binned running averages for 
   the thick- and thin-disc components from GES iDR5 data \citep[blue and red 
     dots, respectively, the vertical bars represent the standard 
     deviations;][]{2020ApJ...888...55F}.

   In the [C/O]--[O/H] plane, a strikingly good agreement is found between the 
   low-Ia/high-Ia sequences of \citet{2019ApJ...886...84G} and the predictions 
   of the thick-disc/thin-disc models featuring rotating massive stars, with 
   the exception of the high-metallicity ends of the tracks, where the models 
   do not match the sudden increase in [C/O] displayed by the observations 
   (Fig.~\ref{fig:fig3}, right panel). \citet{2020ApJ...888...55F} do not 
   provide oxygen abundances for their sample stars, therefore, we cannot 
   compare our theoretical trends to GES data in a [C/O] versus [O/H] diagram. 
   When iron is used as a metallicity tracer (Fig.~\ref{fig:fig3}, left panel), 
   the agreement with the data is much weaker. Although systematic offsets may 
   be present, it is clear that the models completely fail to reproduce the 
   flattening of [C/Fe] versus [Fe/H] for [Fe/H]~$> -0.2$--0.0. This matches 
   the absence of an increase in the theoretical curves for [C/O] versus [O/H] 
   at the high-metallicity end and might be due to underestimated C yields from 
   either low-mass or massive stars, or both. In particular, keeping using 
   solar-metallicity yields for massive stars when the metallicity exceeds 
   solar could lead to underestimate C pollution from Wolf-Rayet stars 
   \citep[see][for a discussion of the effects of metallicity on C production 
     from massive stars]{1986ARA&A..24..329C,1992A&A...264..105M}.

   This section wraps up with a couple of comments on the evolution of the 
   [C/Mg] ratio as a function of [Mg/H]. Magnesium is relatively easy to 
   observe in stars and, thus, it is often used in place of oxygen as a typical 
   $\alpha$-element or metallicity tracer. However, its origin is far less 
   understood than that of oxygen. As a matter of fact, GCE models generally 
   fail to reproduce Mg evolution \citep[see][and references 
     therein]{2010A&A...522A..32R,2018MNRAS.476.3432P} unless some 
   \emph{ad hoc} corrections are made to the yields 
   \citep[e.g.,][]{2004A&A...421..613F}. The yields presented by 
   \citet{2006ApJ...653.1145K} and \citet{2013ARA&A..51..457N} constitute a 
   notable exception, in that they provide a satisfactory fit to the data in 
   the [Mg/Fe]--[Fe/H] plane without any need for adjustments \citep[see, e.g., 
     Fig.~10 of][]{2013ARA&A..51..457N}.

   In Fig.~\ref{fig:fig4} we display the predictions of the parallel model 
   implementing C and Mg yields for massive stars by 
   \citet{2013ARA&A..51..457N}. The low-Ia/high-Ia sequences by 
   \citet{2019ApJ...886...84G} are fitted only marginally, while the almost 
   flat trends of \citet{2020ApJ...888...55F} remain unexplained. We conclude 
   that any meaningful comparison between our GCE model predictions and 
   observations should better involve oxygen rather than magnesium. For this 
   reason, in the remainder of this paper we comment only on C, O, and Fe 
   abundance determinations.

   \subsection{The Galactic bulge}

   It has been pointed out elsewhere \citep{2007ApJ...661.1152F,
     2009A&A...505..605C} that C abundance measurements in the Galactic bulge 
   are extremely important in that they provide a test for the predicted 
   increase in the carbon yield due to enhanced mass loss through stellar winds 
   from metal-rich massive stars.

   Actually, now a non-negligible number of metal-rich stars is found in the 
   solar neighbourhood, which provides a sensible testbed for this idea (see 
   Sect.~\ref{sec:sv} and Figs.~\ref{fig:fig1} to \ref{fig:fig3}). Adding 
   information from bulge stars, however, permits to study not only the 
   dependence on stellar metallicity, but also that on the environment. In 
   fact, by comparing C abundances measured in dwarf stars with similarly high 
   metal content that belong to either the bulge or the local disc, we can 
   contrast the properties of metal-rich stars that formed either fast in a 
   spheroid at high redshift, or at a slower pace and more recently in a 
   flattened structure.

   With [O/H]~= $0.4 \pm 0.28$, $0.4 \pm 0.21$, and $0.62 \pm 0.21$, the three 
   microlensed bulge dwarfs observed at high resolution by 
   \citet{2008ApJ...685..508J} and \citet{2009ApJ...699...66C} probe the 
   high-metallicity tail of the metallicity distribution function of bulge 
   stars. The [C/Fe] ratios determined for these stars lie in the range 
   $\sim$0.0--0.3, while the [C/O] ratios are consistent with solar. Chemical 
   evolution models for the bulge \citep{2019MNRAS.487.5363M} that adopt the 
   same nucleosynthesis prescriptions as models MWG-05 and MWG-06 of 
   \citet{2019MNRAS.490.2838R}, namely, the yields by 
   \citet{2013MNRAS.431.3642V} for low- and intermediate-mass stars and those 
   by \citet{2018ApJS..237...13L} for fast-rotating massive stars (with 
   $\vel_{\rm{rot}} =$ 300 and 150~km~s$^{-1}$, respectively), match very well the 
   [C/Fe] and [C/O] bulge data; on the contrary, the adoption of non-rotating 
   stellar yields leads to severely underestimate (overestimate) the [C/Fe] 
   ([C/O]) ratio (see Fig.~\ref{fig:fig5}).

   More C abundances of dwarf stars, free from stellar evolutionary effects, in 
   the bulge are urgently needed before we can draw any firm conclusion about 
   carbon evolution in the central regions of the Galaxy.

   \subsection{Massive elliptical galaxies}

%-------------------------------------- One column figure
   \begin{figure}
     \centering
     \includegraphics[width=0.48\textwidth]{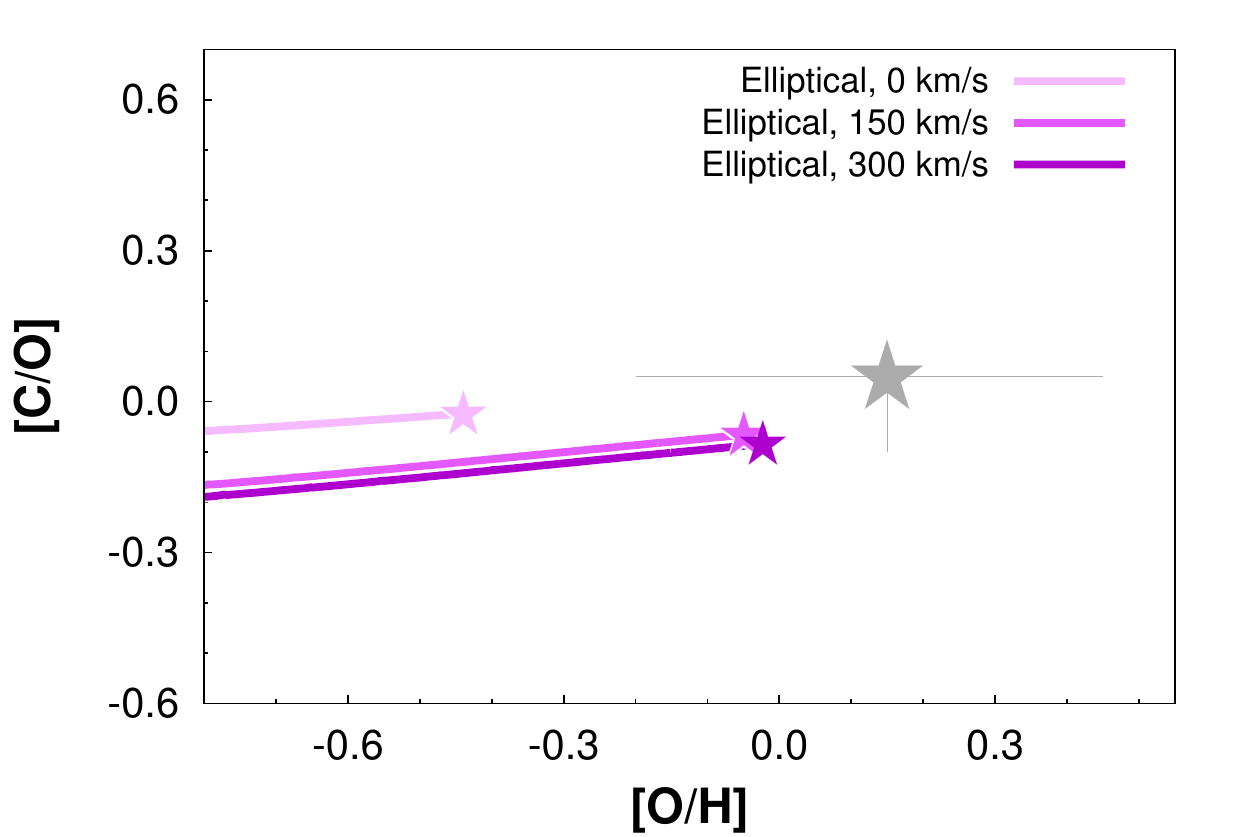}
     \caption{ Average carbon-to-oxygen abundance ratio as a function of [O/H] 
       in the stellar population of the prototype massive elliptical galaxy. 
       The thick solid lines show the evolution of the mass-averaged abundances 
       of the stellar population. The starlets mark the ending points of the 
       theoretical tracks. Different shades of purple refer to different 
       nucleosynthesis prescriptions (see text). The big grey star indicates 
       the values of the ratios estimated from absorption line indices of local 
       galaxies \citep{2012MNRAS.421.1908J}.}
     \label{fig:fig6}
   \end{figure}
%-----------------------------------------------------------------

   In Fig.~\ref{fig:fig6}, we display the evolution of the mean 
   carbon-to-oxygen abundance ratio as a function of [O/H] in the stellar 
   population of the prototype massive elliptical galaxy (thick solid lines, 
   different shades of purple refer to different nucleosynthesis prescriptions, 
   corresponding to different initial rotational velocities of the 
   core-collapse SN progenitors). The galaxy is assumed to form stars for about 
   1~Gyr through a sequence of starbursts, lasting 50~Myr each, alternating 
   with similarly long quiescent periods\footnote{The results do not change if 
     a continuous star formation history of similar duration is assumed 
     instead.} \citep[see][]{2019MNRAS.490.2838R}. The evolutionary tracks end 
   at the time in which a powerful outflow stops the star formation, that is, 
   when the system has reached a stellar mass of about 
   $2 \times 10^{11}$~M$_\odot$. With our choice of the star formation history, 
   the typical massive early-type galaxy is seen to form stars at rates of 
   several hundreds of solar masses per year at redshifts 2 to 3, in agreement 
   with observations of submillimeter galaxies 
   \citep[e.g.,][]{2011MNRAS.412.1913I}.

   The big grey star with generous error bars indicates the region of the 
   [C/O]--[O/H] diagram where passively-evolving, massive SDSS early-type 
   galaxies are found \citep{2012MNRAS.421.1908J}. Observational estimates of 
   the abundance ratios rely on index measurements of unresolved stellar 
   populations. In Fig.~\ref{fig:fig6} we display mean theoretical stellar 
   abundances averaged on the stellar mass. A more precise comparison should 
   involve abundances averaged on the visual light. However, for massive 
   galaxies formed at high redshifts the latter do not differ significantly 
   from the mass-averaged ones \citep[see, e.g.,][and references 
     therein]{2002MNRAS.334..444R,2012ceg..book.....M}. The ending points of 
   the tracks, i.e., the quantities that have to be compared with the 
   observations, are highlighted by the starlets. It is seen that models in 
   which the majority of massive stars are rotating fast comply better with the 
   observations.

   It is worth emphasising at this point that our models for elliptical 
   galaxies adopt a top-heavy gwIMF (see Sect.~\ref{sec:bas}). The adoption of 
   a canonical gwIMF would lead to significantly lower mean stellar 
   metallicities ([O/H]~$\simeq -0.7$ for the models with fast rotators). Thus, 
   the adoption of a gwIMF biased towards massive stars appears to be a 
   fundamental assumption in order to reproduce the data.

   \section{Discussion}
   \label{sec:disc}

   While important uncertainties still plague stellar evolution and 
   nucleosynthesis studies, the flood of data secured by large modern 
   spectroscopic surveys makes it possible, in principle, to define tight 
   abundance trends that can be used to constrain sensibly both stellar and 
   galactic chemical evolution models. In practice, poorly-understood, possibly 
   significant systematics currently hamper our ability to derive accurate 
   trends of abundance ratios over the full metallicity range, even for local 
   samples.

   We have shown that the observational scatter of individual points in the 
   [C/Fe] versus [Fe/H] and [C/Mg] versus [Mg/H] diagrams and the differences 
   in the trends found by the two large surveys considered in this study (GES 
   and GALAH) make it challenging the comparison of the observations with the 
   predictions of GCE models. While individual scattered data points can be 
   fitted reasonably well, the comparison between mean observed and theoretical 
   trends gives less sound results. However, if the scatters are intrinsic, as 
   suggested in Sect.~\ref{sec:hr}, the observational trends should be taken 
   with caution, since they may be affected by biases depending on the nature, 
   origin, and number of individual points used to build them. Moreover, in 
   such a case the usage of mean trends may mask the origin and sources of the 
   scatter. Yet, GCE models, by construction, predict the climate, not the 
   weather (they only account for large-scale phenomena); much more complex 
   (and computationally expensive) hydrodynamical simulations are needed to 
   address specifically the scatter in the data \citep[see, 
     e.g.,][]{2019arXiv190904695E}, but this is beyond the scope of the present 
   paper.

   Regarding the [C/O]--[O/H] diagram for local disc stars, a strikingly good 
   agreement is found between the predictions of our GCE models and targeted 
   observations of thick- and thin-disc stars, as well as the average trends 
   derived from the analysis of GALAH DR2 data \citep{2019ApJ...886...84G}. 

   Overall, our GCE models reproduce satisfactorily well the high resolution C, 
   O, and Fe data for solar neighbourhood stars under the assumption that 
   massive stars rotate fast at low metallicities, while they rotate much 
   slowly, or not at all, for [Fe/H]~$> -1.0$. The bulge and elliptical galaxy 
   data are better matched by models with only fast rotators. This hints to a 
   dependence of stellar rotation on environmental factors, such as pressure, 
   temperature and density of the ambient medium, and/or the effects of the 
   permeating radiation field \citep[see also][]{2019MNRAS.490.2838R}.

   In the framework of our models, more than 60 per cent of the solar C 
   abundance comes from massive stars. This percentage increases, becoming 
   $\sim$70 per cent, for stars in spheroids. Therefore, according to our 
   calculations the majority of C in the Universe comes from massive (fast) 
   rotators, with a non-negligible contribution from intermediate-mass stars. 
   In the Galactic bulge and massive elliptical galaxies C production from 
   high-mass stars is boosted by the adoption of gwIMFs flatter than the 
   canonical one in the high-mass domain. In general, top-heavy gwIMFs are 
   required to explain other observed properties of spheroids 
   \citep[see][]{2012ceg..book.....M} and naturally emerge from the IGIMF 
   theory \citep[][and references 
     therein]{2018A&A...620A..39J,2019A&A...629A..93Y}.

   Finally, it is worth emphasising that, unlike other studies in the 
   literature, we have not applied \emph{ad hoc} corrections to the adopted 
   stellar yield tables to make the model predictions match the observed 
   trends. Yet, we are able to obtain a satisfactory agreement between our 
   model predictions and the observations (when taking into account all the 
   possible sources of uncertainties). That is very encouraging, as it means 
   that C production in stars is now \emph{quantitatively} pretty well 
   understood.

   \section{Conclusions}
   \label{sec:conc}

   In this paper, we present GCE model results for different Galactic 
   components, namely the Galactic thick and thin discs and the bulge, as well 
   for passively-evolving, massive early-type galaxies. In particular, we deal 
   with the evolution of carbon, as manifested in the [C/Fe] versus [Fe/H] and 
   [C/O] versus [O/H] planes (Mg is a less reliable metallicity tracer from the 
   theoretician point of view).

   We conclude that:
   \begin{enumerate}
     \item GCE models fit reasonably well individual data points from targeted 
       high-resolution studies of nearby unevolved stars that belong to the 
       inner halo, thick- and thin-disc components. In particular, both the 
       revised two-infall model recently proposed by 
       \citet{2019A&A...623A..60S} and the parallel model by 
       \citet{2017MNRAS.472.3637G} fit very nicely the data in the [C/O]--[O/H] 
       diagram.
     \item While the models fit very well the [C/O] versus [O/H] trends derived 
       from GALAH DR2 data, several problems are apparent when trying to match 
       the GES and GALAH trends in the [C/Fe]--[Fe/H] and [C/Mg]--[Mg/H] 
       planes. This is partly due to the more uncertain yields of Fe. However, 
       it must also be kept in mind that possibly significant systematics might 
       hamper our ability to derive accurate trends of abundance ratios, even 
       for local stars.
     \item The use of Mg as a metallicity tracer leads to the worst match 
       between theoretical predictions and data, partly due to the poorly-known 
       nucleosynthesis of Mg in stars.
     \item The assumption that most massive stars rotate fast during the 
       earliest phases of galactic evolution leads to the best agreement with 
       the observations for both the Galactic discs, bulge and massive 
       ellipticals.
     \item In order to reproduce the observations, it seems unavoidable that 
       the gwIMF in spheroids is biased towards massive stars. The flattest 
       slopes are expected in the most massive ellipticals. This conclusion 
       strengthens previous claims about the need for a top-heavy gwIMF in 
       submillimeter galaxies based on measurements of $^{13}$C/$^{18}$O 
       isotopic ratios in the gas phase \citep{2018Natur.558..260Z}.
   \end{enumerate}

\begin{acknowledgements}
   This work benefited from the International Space Science Institute (ISSI) in 
   Bern, CH, and the International Space Science Institute-Beijing (ISSI-BJ) in 
   Beijing, CN, thanks to the funding of the team \emph{``Chemical abundances 
     in the ISM: the litmus test of stellar IMF variations in galaxies across 
     cosmic time''} (PIs D.~Romano, Z.-Y. Zhang).
\end{acknowledgements}

\bibliographystyle{aa} % style aa.bst
\bibliography{/Users/donatella/Papers/pap-carbon/R20_carbon_BIB}

\end{document}